\documentclass[twocolumn,dvipdfmx]{aastex61}

\newcommand{\HI}{\text{H\hspace{0.15em}\textsc{i}}}
\newcommand{\HIs}{\text{\scriptsize H\hspace{0.105em}\textsc{i}}}
\newcommand{\NHI}{N_{\HIs}}
\newcommand{\NHIstar}{N_{\HIs}^{\ast}}

\newcommand{\WHI}{W_{\HIs}}

\newcommand{\MHI}{M_{\HIs}}
\newcommand{\tauHI}{\tau_{\HIs}}

\newcommand{\Ts}{T_{\text{s}}}
\newcommand{\XHI}{X_{\HIs}}
\newcommand{\THI}{T_{\HIs}}
\newcommand{\dVHI}{\mathit{\Delta}V_{\HIs}}
\newcommand{\HIcm}{21\,\text{cm}}

\newcommand{\Htwo}{\text{H}_{2}}
\newcommand{\Htwos}{\text{{\scriptsize H}}_{2}}
\newcommand{\NHtwo}{N_{\Htwos}}
\newcommand{\MHtwo}{M_{\Htwos}}

\newcommand{\HII}{\text{H\hspace{0.15em}\textsc{ii}}}
\newcommand{\Halpha}{\text{H}\alpha}

\newcommand{\NH}{N_{\text{H}}}
\newcommand{\MH}{M_{\text{H}}}
\newcommand{\NHref}{N_{\text{H,ref}}}

\newcommand{\CO}{\text{CO}}
\newcommand{\COs}{\text{\scriptsize CO}}
\newcommand{\XCO}{X_{\COs}}
\newcommand{\WCO}{W_{\COs}}

\newcommand{\taud}{\tau_{\text{353}}}
\newcommand{\taudref}{\tau_{\text{353,ref}}}

\newcommand{\Td}{T_{\text{d}}}

\newcommand{\AV}{A_{V}}
\newcommand{\AJ}{A_{J}}
\newcommand{\Ihundred}{I_{\text{100}}}
\newcommand{\Tbg}{T_{\text{bg}}}

\newcommand{\VLSR}{V_{\text{LSR}}}

\newcommand{\Msol}{M_{\odot}}

\newcommand{\Umicron}{\mu\text{m}}
\newcommand{\Umm}{\text{mm}}
\newcommand{\Ucm}{\text{cm}}
\newcommand{\Um}{\text{m}}
\newcommand{\Ukm}{\text{km}}

\newcommand{\Upc}{\text{pc}}

\newcommand{\Us}{\text{s}}

\newcommand{\UGHz}{\text{GHz}}
\newcommand{\UK}{\text{K}}

\newcommand{\Udeg}{\text{deg}}
\newcommand{\Uarcmin}{\text{arcmin}}

\newcommand{\UR}{\text{R}}

\newcommand{\Umag}{\text{mag}}

\newcommand{\UVel}{\Ukm\,\Us^{-1}}
\newcommand{\UII}{\UK\,\UVel} % Velocity Integrated Intensity
\newcommand{\UVND}{\Ucm^{-3}} % Volume Number Density
\newcommand{\UCND}{\Ucm^{-2}} % Column Number Density
\newcommand{\UXHI}{\Ucm^{-2}\,\UK^{-1}\,\Ukm^{-1}\,\Us}
\newcommand{\UXCO}{\Ucm^{-2}\,\UK^{-1}\,\Ukm^{-1}\,\Us}

\newcommand{\MBM}{MBM\,53,\,54,\,55}
\newcommand{\HLCG}{HLC\,G92{$-$}35}
\newcommand{\Planck}{\textit{Planck}}
\newcommand{\IRAS}{\textit{IRAS}}

\newcommand{\RA}{\alpha_{\text{J2000}}}
\newcommand{\Dec}{\delta_{\text{J2000}}}

\received{2016 October 28}
\revised{2017 March 14}
\accepted{2017 March 14}
\submitjournal{ApJ}

\shorttitle{$\HI$, $\CO$, and Dust in the Perseus Cloud}
\shortauthors{Okamoto, R., et al.}
\begin{document}

\title{$\HI$, $\CO$, and Dust in the Perseus Cloud}

\correspondingauthor{Ryuji Okamoto}
\email{okamoto@a.phys.nagoya-u.ac.jp}

\author{Ryuji Okamoto}
\affiliation{Department of Physics, Nagoya University, Chikusa-ku, Nagoya 464-8602, Japan}

\author{Hiroaki Yamamoto}
\affiliation{Department of Physics, Nagoya University, Chikusa-ku, Nagoya 464-8602, Japan}

\author{Kengo Tachihara}
\affiliation{Department of Physics, Nagoya University, Chikusa-ku, Nagoya 464-8602, Japan}

\author{Takahiro Hayakawa}
\affiliation{Department of Physics, Nagoya University, Chikusa-ku, Nagoya 464-8602, Japan}

\author{Katsuhiro Hayashi}
\affiliation{Department of Physics, Nagoya University, Chikusa-ku, Nagoya 464-8602, Japan}

\author{Yasuo Fukui}
\affiliation{Department of Physics, Nagoya University, Chikusa-ku, Nagoya 464-8602, Japan}



\begin{abstract}
Comparison analyses between the gas emission data ($\HI$ $\HIcm$ line and $\CO$ $2.6\,\Umm$ line) and the {\Planck}/{\IRAS} dust emission data (optical depth at $353\,\UGHz$ $\taud$ and dust temperature $\Td$) allow us to estimate the amount and distribution of the hydrogen gas more accurately, and our previous studies revealed the existence of a large amount of optically-thick $\HI$ gas in the solar neighborhood. Referring to this, we discuss the neutral hydrogen gas around the Perseus cloud in the present paper. By using the $J$-band extinction data, we found that $\taud$ increases as a function of the 1.3-th power of column number density of the total hydrogen ($\NH$), and this implies dust evolution in high density regions. This calibrated $\taud\text{--}\NH$ relationship shows that the amount of the $\HI$ gas can be underestimated to be ${\sim}60\%$ if the optically-thin $\HI$ method is used. Based on this relationship, we calculated optical depth of the $\HIcm$ line ($\tauHI$), and found that $\langle\tauHI\rangle\sim0.92$ around the molecular cloud. The effect of $\tauHI$ is still significant even if we take into account the dust evolution. We also estimated a spatial distribution of the $\CO$-to-$\Htwo$ conversion factor ($\XCO$), and we found its average value is $\langle\XCO\rangle\sim1.0\times10^{20}\,\UXCO$. Although these results are inconsistent with some previous studies, these discrepancies can be well explained by the difference of the data and analyses methods.
\end{abstract}




\section{Introduction}

The neutral atomic hydrogen ($\HI$) $\HIcm$ emission was discovered in the Galaxy in 1951, and has been used in order to investigate the structure of the Galaxy. It was revealed that the Galaxy has a spiral structure and is rotating at a nearly uniform velocity of ${\sim}200\,\UVel$ except for within ${\sim}2\,\Upc$ from its center \citep[e.g.,][]{2013PASJ...65..118S}. The emission also brought advances in our understanding of the interstellar medium (ISM) and external galaxies. The role of hydrogen as the raw material of stars also attracted attention. In 1970's, the interstellar carbon monoxide ($\CO$) emission at $2.6\,\Umm$ was discovered and has been used as a tracer of molecular hydrogen ($\Htwo$) clouds, and a picture of the  phase transition from $\HI$ to $\Htwo$ which triggers the star formation emerged. The molecular emission lines in the $\Htwo$ gas is efficient in cooling the gas which leads to release of the cloud internal energy and subsequent gravitational collapse.

Hydrogen (either atomic or molecular) accounts for the majority of the mass of the ISM; in the Galaxy, helium, which is abundant next to hydrogen, and heavier atoms account for ${\sim}25\%$ and ${<}1\%$ in mass, respectively. The volume number densities of the $\HI$ gas and the $\Htwo$ gas are estimated to be ${\sim}1\,\UVND$ and ${\sim}1000\,\UVND$, respectively, on average. The physical states of the neutral gas at the intermediate density around $100\,\UVND$ are not understood into detail. In order to elucidate the evolution of the galaxies, it is an important task to better understand the behavior of the hydrogen gas including the transition from $\HI$ to $\Htwo$ over a density range $1{\text{--}1000}\,\UVND$.

{\Planck} is an astronomical satellite which aimed at observing the cosmic microwave background (CMB), and it observed the all sky at millimeter/sub-millimeter wavelengths \citep[e.g.,][]{2011A&A...536A...1P}. The data obtained by {\Planck} necessarily include the emission from the ISM of the Galaxy as the foreground component of the CMB. By using the {\Planck} data, physical parameters of the interstellar dust such as the optical depth at $353\,\UGHz$ ($\taud$), the dust temperature ($\Td$), etc.\ are obtained, and they are available in the archival form \citep{2014AaA...571A..11P}. These dust parameter data have relative uncertainties of ${\lesssim}10\%$, and hence, we can expect that they accurately reflect the properties and states of the ISM at an angular resolution of $5\,\Uarcmin$. In particular, $\taud$ is much less than $1$ for all over the sky, even toward the Galactic plane, and therefore the data offer a reliable tracer of the column density of the total hydrogen atom ($\NH$), if the dust-to-gas ratio (DGR) is a constant.

\citet{2014ApJ...796...59F,2015ApJ...798....6F} compared the {\Planck} dust data and the gas emission data such as $\HI$ and $\CO$ for the solar neighborhood. The $\HI$ $\HIcm$ emission is generally assumed to be optically thin as written in text books. If the dust properties are uniform and DGR is a constant, it is expected that the velocity-integrated intensity of the $\HIcm$ spectrum ($\WHI$) is proportional to $\taud$ for the data points where the $\CO$ emission is not detected. \citet{2014ApJ...796...59F,2015ApJ...798....6F} however found that the correlation between them is not so good. By introducing $\Td$ into the $\taud${--}$\WHI$ correlation plot, these authors discovered that the poor correlation in the $\taud${--}$\WHI$ plot is mainly due to the data points where the density is high and $\Td$ is low. There are two main possibilities to explain this bad correlation between $\taud$ and $\WHI$; one is the presence of optically-thick $\HI$ gas, and the other is the presence of ``$\CO$-dark $\Htwo$ gas'', which is $\Htwo$ gas without the $\CO$ emission \citep[e.g.,][]{2010ApJ...716.1191W,2011A&A...536A..19P,2014A&A...561A.122L}. \citet{2014ApJ...796...59F,2015ApJ...798....6F} investigated the $\Htwo$ fractions in the hydrogen gas by referring to the UV measurements \citep{2006ApJ...636..891G}, and found that the fractions are typically ${\lesssim}10\%$ toward the lines of sight whose column densities are up to at least $10^{21}\,\UCND$\footnote{Note that in Figure 16 of \citet{2015ApJ...798....6F}, the $\Htwo$ fractions are somewhat larger than $10\%$ toward two Galactic B-type stars, HD 210121 and HD 102065 \citep{2002ApJ...577..221R}. These two stars may be contaminated by their own localized gas, and therefore, the $\Htwo$ fractions for the local ISM are possibly not reliable toward them. For this reason, \citet{2017arXiv170107129F} (see Section~\ref{subsec:tauHI}) did not use the results obtained in \citet{2002ApJ...577..221R}.}. That is, $\HI$ dominates $\Htwo$, and the ``$\CO$-dark $\Htwo$ gas'' would not be a dominant component in the local ISM. Therefore, these authors concluded that there exists a large amount of optically thick $\HI$ gas in the local ISM whose typical optical depth is ${\sim}1$, and the amount of the $\HI$ gas is underestimated by ${\sim}50\%$ if a correction for the opacity effect is not applied. In \citet{2014ApJ...796...59F,2015ApJ...798....6F}, these analyses were made for the high Galactic latitude at $|b|>15^{\circ}$ where the gas density is low. An open issue discussed in \citet{2014ApJ...796...59F,2015ApJ...798....6F} is if $\taud$ obeys a simple linear relationship with the $\NH$ or not. A study in the Orion A molecular cloud \citep{2013ApJ...763...55R} indicates that the dust optical depth is proportional to the $1.28{\text{-th}}$ power of $\NH$ rather than a simple linear relation. If correct, this non-linearity may be ascribed to the dust evolution at high column density. This suggests a possibility to apply a minor modification of the method of \citet{2014ApJ...796...59F,2015ApJ...798....6F} in order to improve the accuracy in $\NH$.

The Perseus cloud is one of the well-known molecular clouds in the solar neighborhood ($d\sim300\,\Upc$) located at $(\ell, b)\sim(160^{\circ}, -20^{\circ})$ \citep[e.g.,][]{2008hsf1.book..308B}. This region is a part of the Gould's belt and the cloud includes some star forming regions (SFRs), IC 348 and NGC 1333, for example. There are a few previous studies on the interstellar hydrogen gas in the Perseus region such as \citet{2012ApJ...748...75L,2014ApJ...784...80L} and \citet{2014ApJ...793..132S}. \citet{2012ApJ...748...75L,2014ApJ...784...80L} estimated a spatial distribution of the $\XCO$ factor, which is a conversion factor between the velocity-integrated intensity of $^{12}\CO(J{=}1{\text{--}}0$) line ($\WCO$) and the column number density of the $\Htwo$ gas ($\NHtwo$). They used the {\IRAS} $60, 100\,\Umicron$ data, and concluded that the average value of $\XCO$ is ${\sim}0.3\times10^{20}\,\UXCO$. This average value is significantly smaller than the typical value for the Galaxy, $(1{\text{--}}2)\times10^{20}\,\UXCO$ \citep[e.g.,][]{2013ARA&A..51..207B}. In \citet{2014ApJ...793..132S}, the measurements of the optical depth of the $\HI$ gas ($\tauHI$) around the Perseus cloud were made. They observed the $\HI$ $\HIcm$ absorption spectra toward 26 extra-Galactic radio continuum sources, and analyzed the data by using the method described in \citet{2003ApJS..145..329H}. They concluded that optically-thick $\HI$ gas was detected only toward ${\sim}15\%$ of the lines of sight, which disagrees with the results of \citet{2014ApJ...796...59F,2015ApJ...798....6F}, and cast doubt on the ``optically-thick $\HI$ gas''.

With these results in mind, the aims of the present study are as follows:
\begin{itemize}
    \item To investigate whether the arguments by \citet{2014ApJ...796...59F,2015ApJ...798....6F} still hold or not in the region where the density is relatively high and there exist the SFRs, and to explore if the $\taud{\text{--}}\NH$ relationship requires the non-linear relationship, suggesting dust evolution. 
    \item To understand the cause for the difference between the previous results by \citet{2014ApJ...796...59F,2015ApJ...798....6F}, and \citet{2012ApJ...748...75L,2014ApJ...784...80L}, \citet{2014ApJ...793..132S} on $\XCO$ and $\tauHI$, and to test the validity of usage of $\taud$ in order to estimate the amount of the hydrogen gas.
\end{itemize}

The present paper is organized as follows. Section~\ref{sec:Observational_Data_set} introduces the data set we used in the present study, and Section~\ref{sec:Analyses_Results} describes the analyses methods and the results. We present the discussions in Section~\ref{sec:Discussions}. Finally Section~\ref{sec:Conclusions} summarizes the paper.

\begin{deluxetable*}{cclc}
\tablewidth{\hsize}
\tablecaption{List of symbols used in the present paper}
\tablenum{1}
\tablehead{\colhead{Symbol} & \colhead{Unit} & \colhead{Description} & \colhead{Note}}
\startdata
$\WHI$     & $[\UK\,\UVel]$   & $\HI$ $\HIcm$ line velocity-integrated intensity.                               & (a)\\
$\NHI$     & $[\UCND]$        & $\HI$ column number density.                                                    & \\
$\NHIstar$ & $[\UCND]$        & $\HI$ column number density (optically thin limit), ${\equiv}\XHI{\times}\WHI$. & \\
$\THI$     & $[\UK]$          & $\HI$ brightness temperature.                                                   & \\
$\dVHI$    & $[\UVel]$        & $\HI$ velocity width, ${\equiv}\WHI/\THI(\text{peak})$.                         & \\
$\tauHI$   & \nodata          & $\HI$ optical depth.                                                            & \\
$\Ts$      & $[\UK]$          & $\HI$ spin temperature.                                                         & \\
$\XHI$     & $[\UXHI]$        & Theoretical conversion factor, ${=}1.823{\times}10^{18}\,\UXHI$.                & \\
$\NHtwo$   & $[\UCND]$        & $\Htwo$ column number density.                                                  & \\
$\NH$      & $[\UCND]$        & Total hydrogen column number density, ${=}\NHI{+}2\,\NHtwo$.                    & \\
$\WCO$     & $[\UK\,\UVel]$   & $^{12}\CO(J{=}1{\text{--}}0)$ line velocity-integrated intensity.               & (b)\\
$\XCO$     & $[\UXCO]$        & Empirical $\CO$-to-$\Htwo$ conversion factor, ${\equiv}\NHtwo/\WCO$.                      & \\
$\taud$    & \nodata          & Optical depth at $353\,\UGHz$ derived by \Planck\slash\IRAS\ satellite.         & \\
$\Td$      & $[\UK]$          & Cold dust temperature derived by \Planck\slash\IRAS\ satellite.                 & \\
$\AJ$      & $[\Umag]$        & $J$-band extinction derived by NICEST method.                                     & (c)\\
$\Tbg$     & $[\UK]$          & Background brightness temperature at $\HIcm$.                                   & (d)\\
$\taudref$ & \nodata          & Reference value of $\taud$, ${=}1.2{\times}10^{-6}$.                            & (e)\\
$\NHref$   & $[\UCND]$        & Reference value of $\NH$, ${=}2.5{\times}10^{20}\,\UCND$.                       & (e)\\
$\alpha$   & \nodata          & Power index in Equation~(\ref{eq:Fukui+2015_eq7}). &
\enddata
%\tablenotetext{}{}
\tablecomments{
    (a) \citet{2011ApJS..194...20P}. The integrated range is from $-188.4\,\UVel$ to $+188.4\,\UVel$.\\
    (b) \citet{2001ApJ...547..792D}. The integrated range is from $-4.9\,\UVel$ to $+12.0\,\UVel$.\\
    (c) \citet{2016AaA...585A..38J}\\
    (d) \citet{1982AaAS...48..219R,1986AaAS...63..205R}\\
    (e) Appendix~\ref{subsec:determination_of_tau353ref_NHref}
    \label{tab:list_symbols}
}
\end{deluxetable*}

\section{Observational Data set}\label{sec:Observational_Data_set}

In the present paper, we used following gas and dust maps in order to investigate relationships between them. These data ware spatially smoothed to an effective HPBW (half power beam width) of $8.4\,\Uarcmin$, which corresponds to that of the CfA $\CO$ data, except for the $\HIcm$ radio continuum map ($35\,\Uarcmin$). Then, they were converted to a $3.0\,\Uarcmin$ grid spacing in Right Ascension and Declination.

\subsection{$\HI$ data}

Data sets of the GALFA $\HI$ survey (Data Release 1) \citep{2011ApJS..194...20P} are used in the present study. This survey was done with the Arecibo Observatory $305\,\Um$ telescope, and its HPBW is ${\sim}4\,\Uarcmin$. Note that we have linearly refitted the baseline of each spectrum using the velocity ranges of mainly $\VLSR\leq-161\,\UVel$ and $\VLSR\geq+112\,\UVel$ in order to correct the baseline offsets in this version of the raw GALFA data. Then, the data were spatially smoothed to be a $8.4\,\Uarcmin$ effective beam size, and RMS fluctuations of the smoothed data are ${\sim}0.06\,\UK$ at a $0.18\,\UVel$ velocity resolution.

\subsection{$\CO$ data}

We used the $^{12}\CO(J{=}1\text{--}0)$ data cube by \citet{2001ApJ...547..792D} as a tracer of $\Htwo$ gas. The HPBW is $8.4\,\Uarcmin$, and the RMS noise fluctuations are ${\sim}0.25\,\UK$ at a $0.65\,\UVel$ velocity resolution \citep{2001ApJ...547..792D}. We downloaded this data cube from the web site of Harvard-Smithsonian Center for Astrophysics (CfA)\footnote{\texttt{https://www.cfa.harvard.edu}}.

\subsection{{\Planck} and {\IRAS} dust emission data}

Archival data sets of dust optical depth at $353\,\UGHz$ ($\taud$) and dust temperature ($\Td$) obtained by \Planck\ and \IRAS\ satellites are used. They were obtained by fitting intensities at $353\,\UGHz$, $545\,\UGHz$, and $857\,\UGHz$ observed with \Planck\ and at $100\,\Umicron$ of IRIS (Improved Reprocessing of the \IRAS\ Survey) with modified-blackbody functions \citep[for details, see][]{2011AaA...536A..24P}. In the present study we utilized version R1.20 of the \Planck\ dust maps\footnote{\texttt{http://irsa.ipac.caltech.edu}} and IRIS $60\,\Umicron$ and $100\,\Umicron$ maps\footnote{\texttt{https://lambda.gsfc.nasa.gov}} \citep{2005ApJS..157..302M}. We downloaded all-sky FITS data sets with a HEALPix\footnote{\texttt{http://healpix.sourceforge.net}} format \citep{2005ApJ...622..759G}.

\subsection{$J$-band extinction data}

An all-sky distribution of $J$-band extinction ($\AJ$) was derived in \citet{2016AaA...585A..38J}. They used ``NICER'' and ``NICEST'' methods in order to calculate the $\AJ$ map based on the 2MASS (two micron all-sky survey) $J$, $H$, and $K$-band extinction data. We downloaded an all-sky FITS data\footnote{\texttt{http://www.interstellarmedium.org}} with a HEALPix format. The angular resolution of the raw map is $3\,\Uarcmin$.

\subsection{$\Halpha$ data}

For the purpose of identifying regions where ultraviolet (UV) emission locally affect the dust properties by heating up or destroying them, we used $\Halpha$ emission data. The $\Halpha$ map was obtained by \citet{2003ApJS..146..407F} and it has an angular resolution of $6\,\Uarcmin$.

\subsection{$\HIcm$ radio continuum data}

$\HIcm$ continuum brightness temperature map is used in order to obtain the background emission. This map was derived by \citet{1986AaAS...63..205R} and includes the $2.7\,\UK$ cosmic background radiation. Although it has a lower spatial resolution (${\sim}35\,\Uarcmin$) than the other maps, we use this map as it is in the present study.

\section{Analyses \& Results}\label{sec:Analyses_Results}

\subsection{Spatial Distributions}\label{subsec:Spatial_Distributions}

\begin{figure*}[p]
    \centering
    \includegraphics[scale=1]{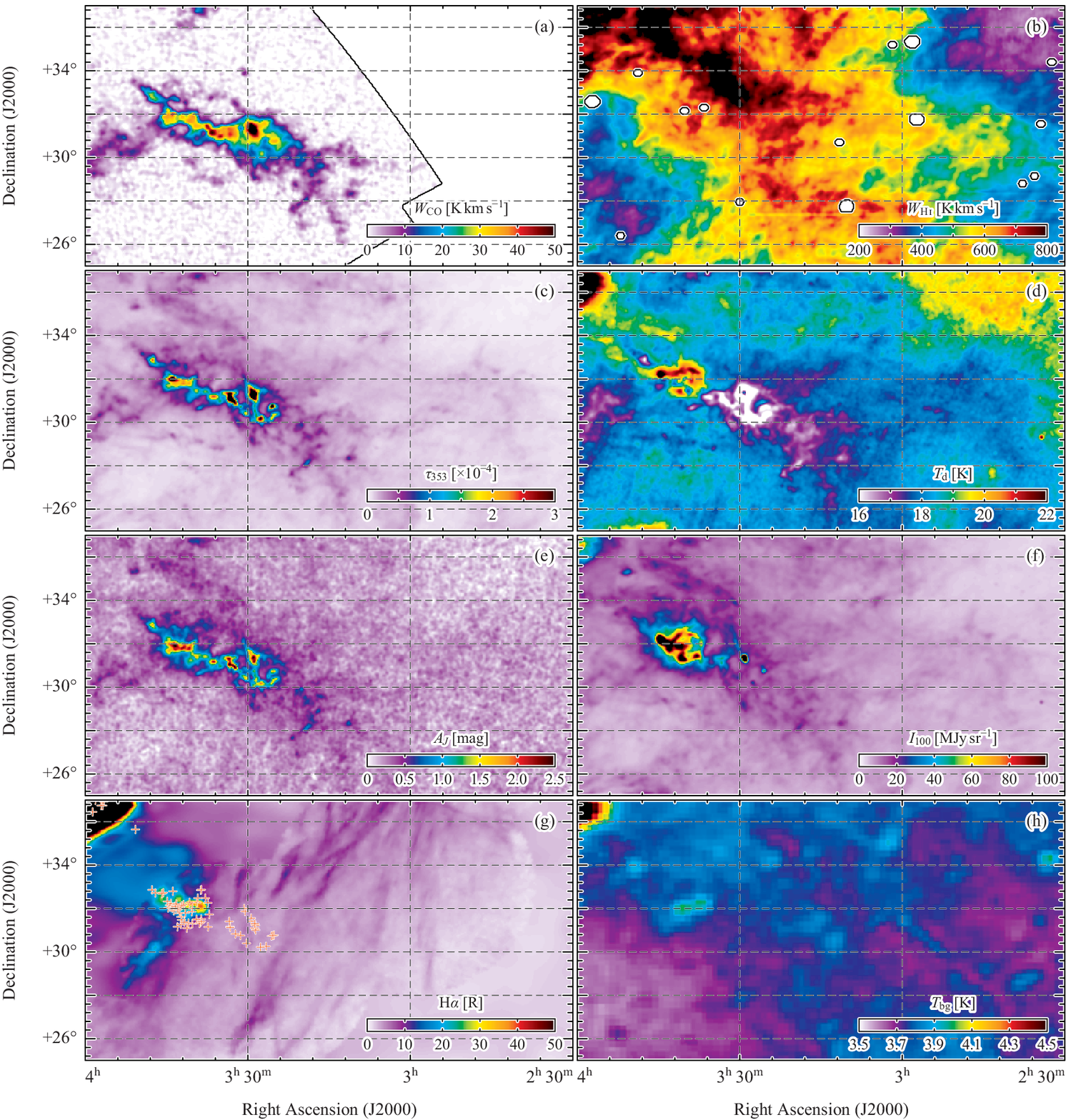}
    \caption{(a) A Velocity-integrated intensity map of $^{12}\CO(J{=}1{-}0)$ line \citep{2001ApJ...547..792D}. The integrated velocity range is from $-4.9\,\UVel$ to $+12.0\,\UVel$. (b) A Velocity-integrated intensity map of $\HI$ $\HIcm$ line \citep{2011ApJS..194...20P}. The integrated velocity range is from $-188.4\,\UVel$ to $+188.4\,\UVel$. See Figure~\ref{fig:Perseus_maps_mask} for the details of the masks. (c) and (d) Spatial distributions of optical depth at $353\,\UGHz$ and cold dust temperature obtained by the {\Planck}/{\IRAS} satellites \citep{2011AaA...536A..24P}. (e) A map of $J$-band extinction derived by 2MASS data and ``NICEST'' analyses \citep{2016AaA...585A..38J}. (f) An intensity map at $100\,\Umicron$ obtained by {\IRAS} satellite \citep{2005ApJS..157..302M}. (g) An $\Halpha$ intensity map derived by \citet{2003ApJS..146..407F}. Red crosses indicate positions of YSO candidates \citep{2014PASJ...66...17T}. (h) A spatial distribution of the $21\,\Ucm$ background radio continuum \citep{1986AaAS...63..205R}. This map includes the $2.7\,\UK$ cosmic background radiation, etc. Effective beam sizes are $8.4\,\Uarcmin$ for the panels (a){--}(g), and $35\,\Uarcmin$ for the panel (h).}
    \label{fig:Perseus_maps_total}
\end{figure*}

Figure~\ref{fig:Perseus_maps_total} shows spatial distributions of data sets used in the present study.

Figure~\ref{fig:Perseus_maps_total}(a) is the distribution of velocity-integrated intensity of $^{12}\CO(J{=}1{-}0)$ line ($\WCO$) obtained by \citet{2001ApJ...547..792D}. The integrated velocity range is from $-4.9\,\UVel$ to $+12.0\,\UVel$. We use this map as a tracer of column number densities of molecular hydrogen ($\NHtwo$).

Figure~\ref{fig:Perseus_maps_total}(b) is a map of velocity-integrated intensity of $\HI$ $\HIcm$ line ($\WHI$). It was derived by GALFA-$\HI$ survey \citep{2011ApJS..194...20P}. The integrated velocity range is from $-188.4\,\UVel$ to $+188.4\,\UVel$. Because we found some strong spike noises in the spectra in our region, we masked 4 areas to exclude these bad spectra. See Figure~\ref{fig:Perseus_maps_mask} for details on the mask.

Figure~\ref{fig:Perseus_maps_total}(c) and (d) show distributions of the dust emission parameters obtained from the {\Planck}/{\IRAS} observations. A map of dust optical depth at $353\,\UGHz$ ($\taud$) is drawn in Figure~\ref{fig:Perseus_maps_total}(c). $\taud$ is obtained by using the data observed with the {\Planck} satellite, which is aimed at accurately detecting the spatial distribution of the CMB. Therefore, it has high reliability with typical uncertainty of ${\lesssim}10\%$. In addition, the value of $\taud$ is at most ${\lesssim}10^{-3}$ even toward the Galactic plane, and hence we can expect that it perfectly reflects the distribution and amount of ISM. $\taud$ is considered as a high accuracy{\slash}precision tracer of column number density of total hydrogen atoms ($\NH$) if DGR is uniform.

Figure~\ref{fig:Perseus_maps_total}(d) shows cold dust temperature ($\Td$) obtained simultaneously with $\taud$. We can see a trend that $\Td$ becomes lower toward the molecular clouds detected by $\CO$. It indicates that   dust grains themselves shield the interstellar radiation field (ISRF) in the molecular clouds and grains are less heated in addition to radiative cooling (see also Figure~\ref{fig:Perseus_corr_tau353_Td}). Note that $\Td$ is high around local heating sources, such as the California Nebula (NGC 1499), the Pleiades (M45), and the ``ring-like feature'' described in \citet{2012ApJ...748...75L}. The $\Td$ image has a relatively low contrast compared to $\taud$, in particular the range of $\Td$ in this region is typically from $17\,\UK$ to $22\,\UK$. This is because $\Td$ is proportional to the one-sixth power of the total intensity radiated from thermal-equilibrium grains expressed as the modified-blackbody model. It reflects that the total radiation energy of the modified-blackbody is proportional to the fourth power of the temperature, and  is in addition proportional to the emissivity spectral index (${\sim}2$).

Figure~\ref{fig:Perseus_maps_total}(e) is an image of $J$-band extinction ($\AJ$) derived by \citep{2016AaA...585A..38J} using 2MASS data and ``NICEST'' method \citep{2001AaA...377.1023L}. We use this map as an another tracer of $\NH$.

A spatial distribution of IRIS $100\,\Umicron$ intensity ($\Ihundred$) is drawn in Figure~\ref{fig:Perseus_maps_total}(f). This map is used for a comparison of the $\CO$-to-$\Htwo$ conversion factor (see Section~\ref{subsec:XCO}).

We show an intensity map of $\Halpha$ emission in Figure~\ref{fig:Perseus_maps_total}(g), overlaid with locations of young stellar object (YSO) candidates. $\Halpha$ emission indicates the presence of ionized hydrogen ($\HII$) and thus indicates the presence of strong UV radiation. We use the data in order to identify the region where the dust may be destroyed by UV radiation. The red crosses are the location of YSO candidates catalogued by \citet{2014PASJ...66...17T}. From this catalogue we extracted the candidates under the condition that ``probability of being a YSO candidate'' is greater than $0.997$. It is based on AKARI Far-Infrared Surveyor (FIS) Bright Source Catalogue.

Figure~\ref{fig:Perseus_maps_total}(h) shows a spatial distribution of brightness temperature of $\HIcm$ continuum radiation. This map was derived by \citet{1986AaAS...63..205R} and includes the $2.7\,\UK$ cosmic background radiation, etc.

\subsection{Masking}\label{subsec:Masking}

\begin{figure*}[]
    \centering
    \includegraphics[scale=1]{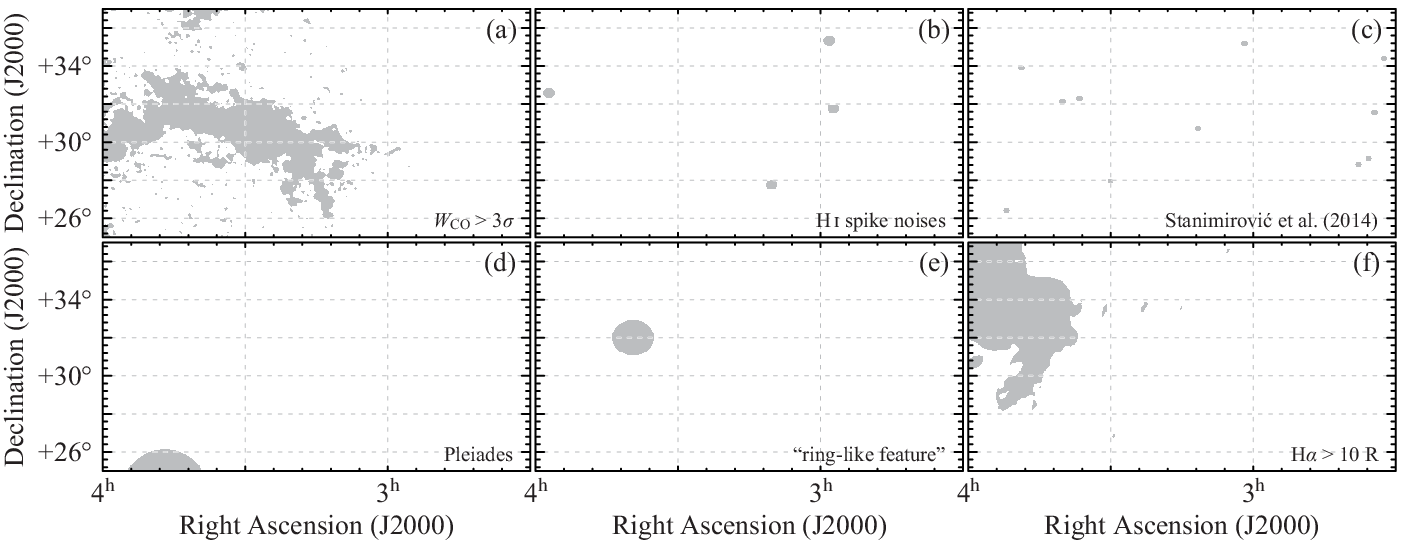}
    \caption{Mask maps used in the present study. We masked the data points (a) where $\WCO\geq3\,\sigma$, (b) where spike noises are detected in $\HI$ spectra, (c) within an effective HPBW ($8.4\,\Uarcmin$) of background radio point sources described in \citet{2014ApJ...793..132S}, (d) within $1.5\,\Udeg$ of Pleiades, (e) within $56\,\Uarcmin$ from $(\RA, \Dec)=(3^{\text{h}}\,39^{\text{m}}\,30^{\text{s}}, +32^{\circ})$ \citep[``ring-like feature'' described in][]{2012ApJ...748...75L}, and (f) where $\Halpha\geq10\,\UR$. (b) and (c) are applied to $\HI$ maps, (d), (e), and (f) are applied to dust maps.}
    \label{fig:Perseus_maps_mask}
\end{figure*}

Masks applied in the present study are shown in Figure~(\ref{fig:Perseus_maps_mask}). 

\begin{enumerate}\renewcommand{\labelenumi}{(\alph{enumi})}
    \item This mask is used in order to mask the data points which molecular hydrogen can exist ($\WCO>3\,\sigma$).
    \item There exist strong spike noises (equivalent to $\THI\sim30\,\UK$) in some spectra of GALFA-$\HI$ DR1 data. This mask is used in order to exclude the data points within $8.4\,\Uarcmin$ ($=1\,\text{HPBW}$) from these bad spectra.
    \item In \citet{2014ApJ...793..132S} absorption spectra due to the local $\HI$ gas are detected toward extra-Galactic radio continuum sources. Since there are 11 out of these sources in this region, this mask is used to exclude the data points within $8.4\,\Uarcmin$ ($=1\,\text{HPBW}$) from them.
    \item This mask excludes the data points within $1.5\,\Udeg$ from the center of the Pleiades $(\RA, \Dec)\sim(3^{\text{h}}\,47^{\text{m}}, +24^{\circ}\,07^{\text{m}})$ \citep[SIMBAD Astronomical Database, ][]{2000AaAS..143....9W} because dust grains are locally heated up.
    \item \citet{2012ApJ...748...75L} describes the ``ring-like feature'', which is formed by an $\HII$ region due to a B-type star named HD 278942 \citep{2006ApJ...643..932R}. As well as \citet{2012ApJ...748...75L} this mask is used to exclude the data points within $56\,\Uarcmin$ from $(\RA, \Dec)=(3^{\text{h}}\,39^{\text{m}}\,30^{\text{s}}, +32^{\circ})$.
    \item Dust grains can be locally heated up or destroyed in regions where $\Halpha$ emission is strong. We use this mask for the purpose of excluding the data points where $\Halpha\geq10\,\UR$.
\end{enumerate}

Mask-(a) is applied to Figure~\ref{fig:Perseus_corr_tau353_WHI}, \ref{fig:Perseus_maps_NHratio}(c), \ref{fig:Perseus_maps_tauHI_Ts}, \ref{fig:Perseus_histograms_tauHI_Ts_NH}. Mask-(b), (c) are applied to the $\HI$ data, and mask-(d), (e), (f) to the dust data.

\subsection{Velocity Structure}\label{subsec:Velocity_Structure}

\begin{figure*}[]
    \centering
    \includegraphics[scale=1]{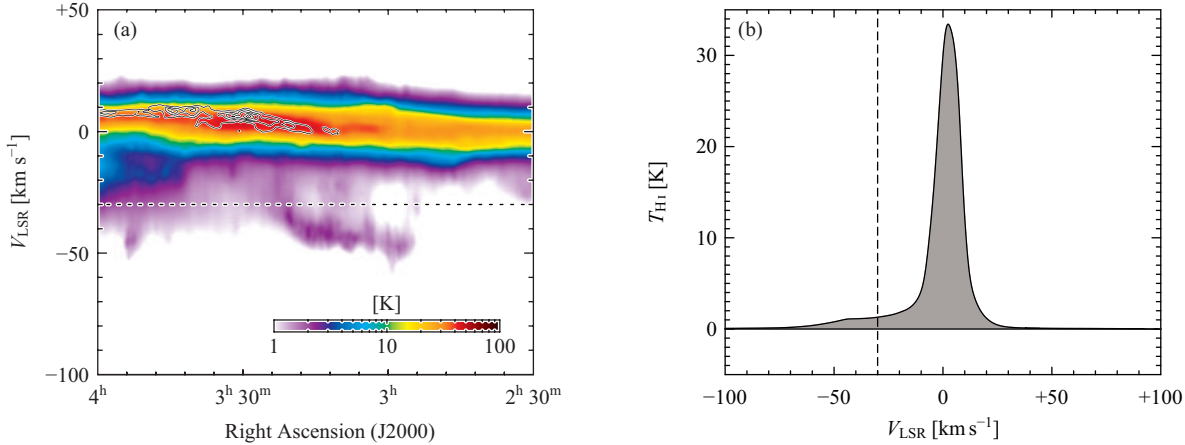}
    \caption{(a) A R.A.-velocity diagram of $\HI$ (image) and $\CO$ (contours) in terms of Dec.-averaged intensities. The contours are drawn every $0.3\,\UK$ from $0.3\,\UK$. The average is calculated over the whole Dec.\ range shown in Figure (\ref{fig:Perseus_maps_total}). (b) An averaged $\HI$ spectrum calculated over the whole region shown in Figure (\ref{fig:Perseus_maps_total}) (b). The dashed lines indicate $\VLSR=-30\,\UVel$.}
    \label{fig:Perseus_PV_meanspectrum}
\end{figure*}

Figure~\ref{fig:Perseus_PV_meanspectrum}(a) is a R.A.-velocity diagram of $\HI$ (image) and $\CO$ (contours, every $0.3\,\UK$ from $0.3\,\UK$) in terms of Dec.-averaged intensities. Figure~\ref{fig:Perseus_PV_meanspectrum}(b) shows the mean $\HI$ spectrum of this region. Although there is a intermediate-velocity component (intermediate-velocity cloud; IVC) at $\VLSR\lesssim -30\,\UVel$, its contribution to the column density is only up to ${\sim}6\%$ of the total, we therefore use the data including all the velocity components.

\subsection{Hydrogen Amount Estimation}\label{Hydrogen_Amount_Estimation}

\begin{figure}[]
    \centering
    \includegraphics[scale=1]{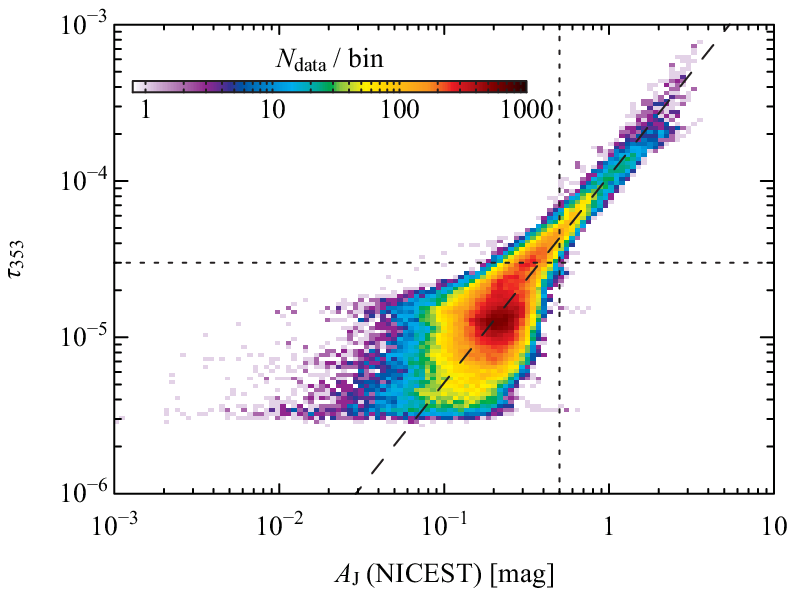}
    \caption{A double logarithmic scatter plot (density plot) between $\AJ$ and $\taud$. Colors are correspond to numbers of the data points included in each bin. The dashed line shows a relationship of $\taud=(1.08\times10^{-4})\times(\AJ)^{1.32}$, which is the result of a linear regression in double logarithmic scale using the data of $\AJ\geq0.5\,\Umag$ and $\taud\geq3\times10^{-5}$. The dotted lines indicate $\AJ=0.5\,\Umag$ and $\taud=3\times10^{-5}$.}
    \label{fig:Perseus_corr_AJ_tau353}
\end{figure}

Figure~(\ref{fig:Perseus_corr_AJ_tau353}) is a double logarithmic correlation plot between $\AJ$ and $\taud$. A linear regression in double logarithmic scale using the data of $\AJ\geq0.5\,\Umag$ and $\taud\geq3\times10^{-5}$ yields the following relationship,
\begin{eqnarray}
    \taud&=[(1.08\pm0.02)\times10^{-4}]\times(\AJ)^{1.32\pm0.04}.
    \label{eq:fit_result_corr_tau353_AJ}
\end{eqnarray}
We assume the uncertainties in the $\AJ$ data are $0.2\,\Umag$, which is the typical fluctuation nearby the Perseus cloud. Considering $\AJ$ as a linear tracer of $\NH$ along each line of sight, Equation~(\ref{eq:fit_result_corr_tau353_AJ}) indicates that dust emission cross section per hydrogen atom increases with the ${\sim}1.3$-th power of $\NH$. A similar analysis was done in \citet{2013ApJ...763...55R} for the Orion A region, and they concluded that the power index is 1.28. Our result is consistent with it, and these results show dust evolution in high density regions. The grain size is typically up to ${\sim}0.2\,\Umicron$ \citep{2013A&A...558A..62J}, which is much smaller than the $J$-band wavelength, ${\sim}1.25\,\Umicron$. In addition, the grain size can change by only ${\sim}0.02\,\Umicron$ \citep{2015A&A...577A.110Y}. \citet{2013ApJ...763...55R} and \citet{2015A&A...580A.114F} also argue that in the high density regions it is mostly the sub-millimeter opacity that is changing, not the infrared extinction. Therefore, changes in the grain size have small effect on the $J$-band extinction, and we can regard $\AJ$ as a linear tracer of $\NH$ if DGR is uniform. In the present study, therefore, we adopted $1.3$ as the power index ($\alpha$) from Equation~(\ref{eq:fit_result_corr_tau353_AJ}).

In \citet{2014ApJ...796...59F,2015ApJ...798....6F} we suggested an algorithm for estimating the amount of the hydrogen gas using $\taud$ as an accurate tracer of $\NH$.  In \citet{2014ApJ...796...59F} we assumed the relationship $\taud\propto\NH$. On the other hand, \citet{2015ApJ...798....6F}, considering the result of \citet{2013ApJ...763...55R}, made discussion using the relationship of
\begin{eqnarray}
    \frac{\taud}{\taudref}=\left(\frac{\NH}{\NHref}\right)^{\alpha}.
    \label{eq:Fukui+2015_eq7}
\end{eqnarray}
Here $\taudref$ and $\NHref$ are normalization constants which satisfy the relationship of $\NHref=(1.15\times10^{8})\times\XHI\times\taudref$ \citep{2015ApJ...798....6F}. In the present study we use Equation~(\ref{eq:Fukui+2015_eq7}) by substituting $\alpha=1.3$, $\taudref=1.2\times10^{-6}$, and $\NHref=2.5\times10^{20}\,\UCND$;
\begin{eqnarray}
    \NH&=\left(\frac{\taud}{\taudref}\right)^{1/\alpha}\,\NHref\label{eq:calc_NH}\\
    &=(9.0\times10^{24})\times(\taud)^{1/1.3}.\nonumber
\end{eqnarray}
See Appendix~\ref{subsec:determination_of_tau353ref_NHref} for the details of the determination of $\taudref$ and $\NHref$.

Conventionally, the following equation is used to calculate the column number density of $\HI$ gas from an observable value, $\WHI$,
\begin{eqnarray}
    \NHIstar\equiv\XHI\times\WHI.
    \label{eq:calc_NHI*}
\end{eqnarray}
Although widely used, this equation is obtained under the assumption that the $\HI$ gas is optically thin ($\tauHI\ll1$). The symbol $\NHIstar$ is the $\HI$ column number density in the optically-thin limit. Therefore, when Equation~(\ref{eq:calc_NHI*}) is used, the $\HI$ column number density can be underestimated if optically-thick $\HI$ gas exists. Here we will show that $\HI$ optical depth has significant effects on the estimation of the $\HI$ amount in the Perseus region.

\begin{figure}[]
    \centering
    \includegraphics[scale=1]{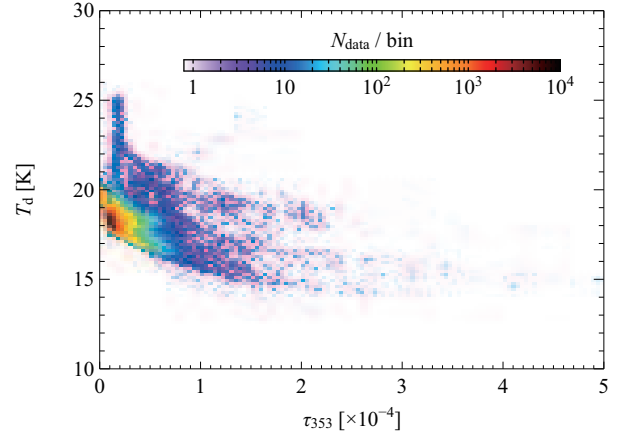}
    \caption{A scatter plot (density plot) between $\taud$ and $\Td$. It clearly shows a negative relationship as with \citet{2014ApJ...796...59F} for MBM 53, 54, 55 region.}
    \label{fig:Perseus_corr_tau353_Td}
\end{figure}

Figure~\ref{fig:Perseus_corr_tau353_Td} is a scatter plot between $\taud$ and $\Td$ of the region analyzed. As well as the MBM 53, 54, 55 region \citep{2014ApJ...796...59F}, it clearly shows an anti-correlation relationship between the two quantities. As described in Section~\ref{subsec:Spatial_Distributions}, dust grains are heated by interstellar radiation field (ISRF). In the areas where $\taud$ is small, the amount of ISM is also small and therefore the ISRF efficiently heats up the dust grains. Conversely, in the areas where $\taud$ is large, the amount of ISM is also large and the ISRF is shielded by the grains themselves, and they are cooled by dust radiation. The anti-correlation in Figure~\ref{fig:Perseus_corr_tau353_Td} reflects these picture. In addition, in the latter case, the dust growth can increase $\taud$ in such high density areas.

\begin{figure*}[]
    \centering
    \includegraphics[scale=1]{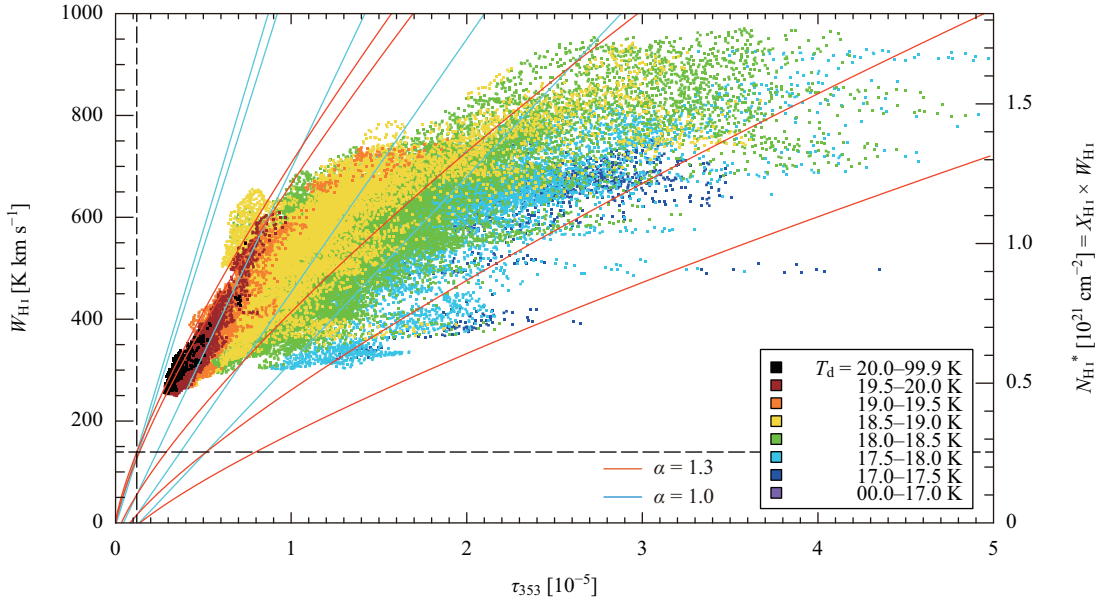}
    \caption{A scatter plot between $\taud$ and $\WHI$ colored by $\Td$ in $0.5\,\UK$ windows. The right hand $y$-axis indicates the column number density of $\HI$ in the optically-thin limit ($\NHI$). The curves are theoretical relationships between both variables (Equation~(\ref{eq:curve_tau353_WHI_theoretical})) for the case of $\tauHI=0, 0.11, 1, 2, 3$ (from left to right). The light blue ones are for the case of $\alpha=1.0$ and red ones for the case of $\alpha=1.3$, respectively. We considered $\Tbg=3.7\,\UK$ and $\dVHI=13.7\,\UVel$ (mean values in this region) as constants. The dashed lines correspond to $\taudref=1.2\times10^{-6}$ and $\NHref=2.5\times10^{20}\,\UCND$ (Appendix~\ref{subsec:determination_of_tau353ref_NHref}).}
    \label{fig:Perseus_corr_tau353_WHI}
\end{figure*}

Figure~\ref{fig:Perseus_corr_tau353_WHI} is a correlation plot between $\taud$ and $\WHI$ colored by $\Td$ in windows of $0.5\,\UK$ intervals. Note that we applied the mask-(a) (CO mask) and therefore, $\NH=\NHI$ for the data points plotted in Figure~\ref{fig:Perseus_corr_tau353_WHI}. Although the correlation is not so good as a whole, the distributions of the data points on the $\taud\text{--}\WHI$ plane are clearly different when separated according to $\Td$. The scattering is small for the data points where $\Td$ is high, and it becomes large with decreasing $\Td$. In particular, the distribution of the high-$\Td$ points (such as $\Td>19.5\,\UK$) is elongated, and it seems to pass through the origin when it is extrapolated. On the other hand, the distribution of the low-$\Td$ points is broadened along the $\taud$ axis. As we described above, this indicates that for such low-$\Td$ points the amount of the ISM is large, and hence $\WHI$ is saturated against $\taud$ because of the effect of large optical depth, i.e., $\tauHI\gtrsim0.3$. Therefore, for such data points the amount of $\HI$ cannot be calculated by using Equation~(\ref{eq:calc_NHI*}). It can be said that the reason of the low spatial correlation between $\taud$ and $\WHI$ (Figure~\ref{fig:Perseus_maps_total}(b), (c)) is the effect of $\tauHI$. Note that there is another possibility to explain the bad correlation between $\taud$ and $\WHI$; the presence of ``$\CO$-dark $\Htwo$ gas'', which is not detectable by the $\CO$ line. In \citet{2015ApJ...798....6F}, however, the authors examined the fraction of $\Htwo$ in the hydrogen gas by referring to the results of the UV measurements \citep{2006ApJ...636..891G,2002ApJ...577..221R}, and found that the fraction is at most ${\sim}10\%$. This means that $\HI$ dominates $\Htwo$ in the typical hydrogen gas. In addition, \citet{2012ApJ...746...82F} revealed that the total hydrogen column number density ($\NHI+2\,\NHtwo$) shows a good correlation with the TeV gamma ray distribution (a reliable tracer of the total hydrogen) in the supernova remnant RX J1713.7$-$3946 when the $\HI$ optical depth is corrected. From these, the bad correlation between $\taud$ and $\WHI$ can be explained by optically-thick $\HI$ gas alone, without ``$\CO$-dark $\Htwo$ gas''. In the present study therefore, we did not consider ``$\CO$-dark $\Htwo$ gas''. The aspect of the $\WHI$ saturation can be explained by using the radiation transfer equation as stated below. 

The following Equations~(\ref{eq:radiation_transfer}) and (\ref{eq:def_tauHI}) are the radiation transfer equation of $\HI$ $\HIcm$ line and the equation of $\HI$ optical depth, respectively, and both of them are derived theoretically \citep[e.g.,][]{2011piim.book.....D,1990ARA&A..28..215D}.
\begin{eqnarray}
    \WHI&=(\Ts-\Tbg)\,\dVHI\,\{1-\exp(-\tauHI)\}\label{eq:radiation_transfer}\\
    \tauHI&=\frac{\NHI}{\XHI}\,\frac{1}{\Ts}\,\frac{1}{\dVHI}\label{eq:def_tauHI}
\end{eqnarray}
These equations are independent of each other, and are valid regardless of whether $\tauHI$ is negligible or not.We define $\dVHI$ as $\WHI/\THI(\text{peak})$, and $\tauHI$ in Equations~(\ref{eq:radiation_transfer}) and (\ref{eq:def_tauHI}) are regarded as the average values over the velocity range $\dVHI$. From the two and Equation~(\ref{eq:calc_NH}) we derive the following relationship;
\begin{eqnarray}
    \WHI=\left\{\left(\frac{\taud}{\taudref}\right)^{1/\alpha}\,\frac{\NHref}{\XHI}\,\frac{1}{\tauHI}\,\frac{1}{\dVHI}-\Tbg\right\}\nonumber\\
    \times\dVHI\,\{1-\exp(-\tauHI)\}.
    \label{eq:curve_tau353_WHI_theoretical}
\end{eqnarray}
The lines/curves in Figure~\ref{fig:Perseus_corr_tau353_WHI} show this relationships in cases of $\alpha=1.0$ and $1.3$ (light blue lines and red curves, respectively) when $\tauHI\ll1, \tauHI=0.11, 1, 2$, and $3$ are substituted (from left to right). We applied $\dVHI=13.7\,\UVel$ and $\Tbg=3.7\,\UK$, which are the mean values of the data points shown in Figure~\ref{fig:Perseus_corr_tau353_WHI}. The standard deviations of $\dVHI$ and $\Tbg$ are $1.9\,\UVel$ and $0.05\,\UK$, respectively. Note that $\tauHI=0.21$ is given in Appendix~\ref{subsec:determination_of_tau353ref_NHref}. According to these curves, there is a trend that larger $\tauHI$ gives a smaller slope, which is consistent with the distribution of the low-$\Td$ points.  Therefore, these equal-$\tauHI$ curves should trace the distribution of the equal-$\Td$ points. As shown in Figure~\ref{fig:Perseus_corr_tau353_WHI}, the curves to which $\alpha=1.3$ is applied show better correlations with the distributions of the data points rather than $\alpha=1.0$ for both high- and low-$\Td$. This also suggests that it is important to take into account the dust evolution. We can say in addition that the small variances at high $\Td$ (small $\tauHI$) suggest the uniform DGR and the uniform grain size.

\begin{figure*}[]
    \centering
    \includegraphics[scale=1]{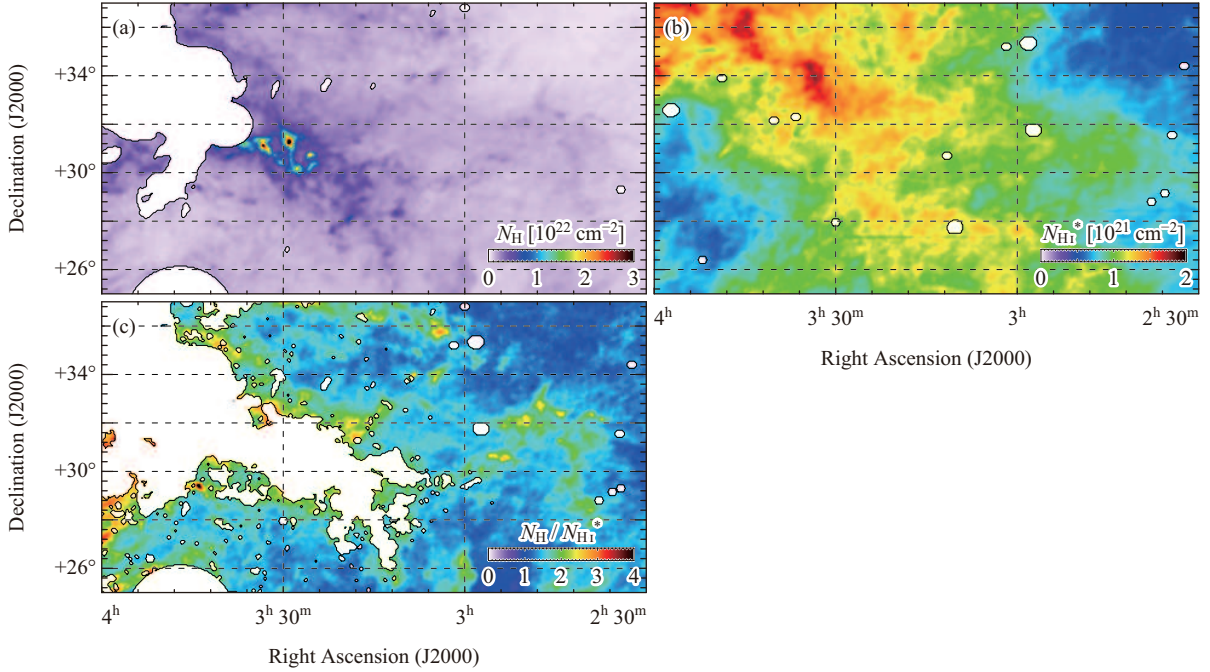}
    \caption{(a) A map of $\NH$ calculated by Equation~(\ref{eq:calc_NH}). Masks-(d), (e), (f) are applied. (b) A spatial distribution of $\HI$ column density calculated under the assumption of the optically thin limit, $\NHIstar$ ($=\XHI\times\WHI$) Masks-(b), (c) are applied. (c) A pixel-to-pixel ratio map between (a) and (b) ($\NH/\NHIstar$). We found the mean value as $\langle\NH/\NHIstar\rangle\sim1.6$. Mask-(a) is applied in addition.}
    \label{fig:Perseus_maps_NHratio}
\end{figure*}

Figure~\ref{fig:Perseus_maps_NHratio}(a) is a spatial distribution of $\NH$ calculated with Equation~(\ref{eq:calc_NH}). Figure~\ref{fig:Perseus_maps_NHratio}(b) shows a distribution of $\NHIstar$ derived with Equation~(\ref{eq:calc_NHI*}), which is the $\HI$ column number density at optically-thin limit. Note that in Figure~\ref{fig:Perseus_maps_NHratio}(a) and (b) the scales of their color-bars are significantly different from each other. In Figure~\ref{fig:Perseus_maps_NHratio}(c) we show a distribution of $\NH/\NHIstar$, which is the pixel-to-pixel ratio map of Figure~\ref{fig:Perseus_maps_NHratio}(a) and (b). In this map, mask-(a) is applied in order to exclude the data points where $\Htwo$ can exist. We found that the mean value of $\NH/\NHIstar$ is ${\sim}1.8$ (see also Figure~\ref{fig:Perseus_histograms_tauHI_Ts_NH}), and hence in this region, the amount of $\HI$ is underestimated to be ${\sim}57\%$ if Equation~(\ref{eq:calc_NHI*}) is used. From these, it turns out that $\HI$ optical depth has a significant effect in estimating the amount of $\HI$ gas in the ISM.

Using Equation~(\ref{eq:calc_NH}), we can estimate the total (atomic and molecular) amount of the hydrogen gas in a high accuracy without being affected by $\HI$ optical depth. However, note that Equation~(\ref{eq:calc_NH}) alone cannot separate atomic and molecular components, and also cannot separate individual components located along the same line of sight.

\subsection{$\tauHI$ and $\Ts$ Estimation}\label{sec:tauHI_Ts_Estimation}

\begin{figure}[]
    \centering
    \includegraphics[scale=1]{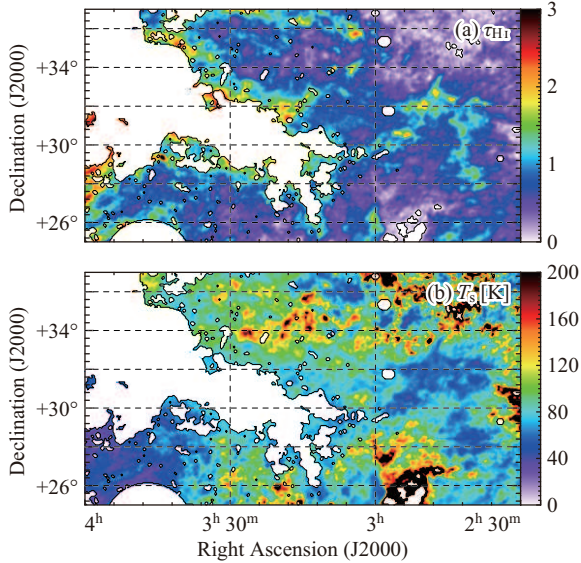}
    \caption{Estimated spatial distributions of (a) $\tauHI$ and (b) $\Ts$, which are the solutions of the system of equations (\ref{eq:radiation_transfer}) and (\ref{eq:def_tauHI}). For the purpose of comparisons with the results of \citet{2014ApJ...793..132S}, we spatially interpolated the $\HI$ spectra toward the background sources using the bilinear interpolation method and computed $\tauHI$ and $\Ts$.}
    \label{fig:Perseus_maps_tauHI_Ts}
\end{figure}

Once $\NH$ is derived with Equation~(\ref{eq:calc_NH}), $\tauHI$ and $\Ts$ can be calculated independently as solutions of a system of equations (\ref{eq:radiation_transfer}) and (\ref{eq:def_tauHI}) \citep{2014ApJ...796...59F,2015ApJ...798....6F}. Since the solutions of the coupled equations cannot be expressed analytically, they are numerically calculated. We note that in the limit of $\tauHI\ll1$ the solutions are indeterminate. Figure~\ref{fig:Perseus_maps_tauHI_Ts} shows the spatial distributions of (a) $\tauHI$ and (b) $\Ts$, respectively. In order to compare with the result of $\tauHI$ obtained based on $\HI$ absorption measurements in \citet{2014ApJ...793..132S}, we spatially interpolated (bilinear method) the $\HI$ spectra masked by mask-(c) and calculated the solutions.

\begin{figure*}[]
    \centering
    \includegraphics[scale=1]{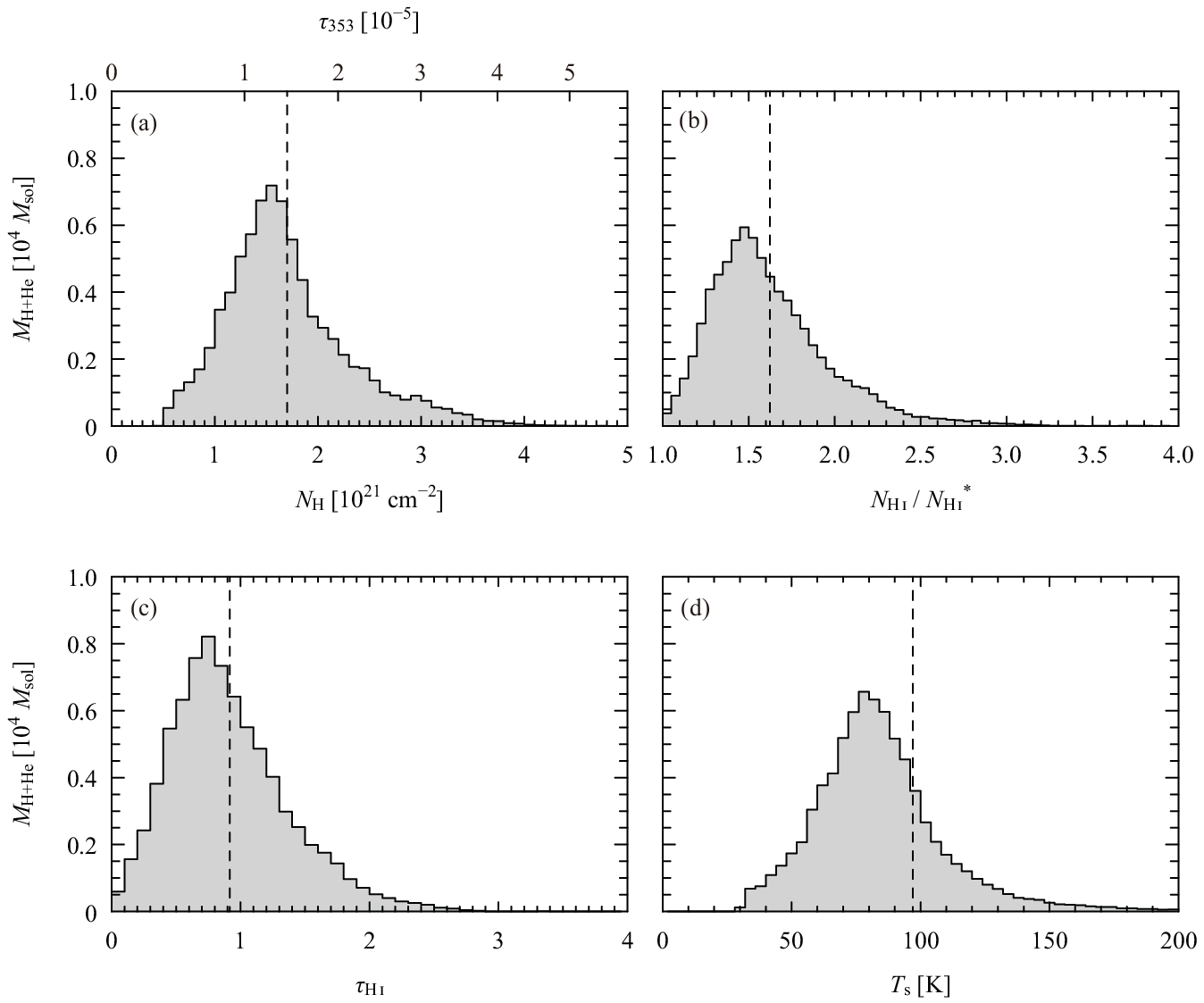}
    \caption{Mass-weighted histograms of (a) $\NH$, (b) $\NHI/\NHIstar$, (c) $\tauHI$, and (d) $\Ts$  for the region shown in Figure~(\ref{fig:Perseus_maps_tauHI_Ts}). We assume a distance of $300\,\Upc$ and the mass includes helium. The dashed lines indicate the mass-weighted mean values of each histogram. In panel (a) the upper $x$-axis shows the $\taud$ calculated by using Equation~(\ref{eq:calc_NH}).}
    \label{fig:Perseus_histograms_tauHI_Ts_NH}
\end{figure*}

In Figure~\ref{fig:Perseus_histograms_tauHI_Ts_NH} we show mass-weighted histograms of (a) $\NH$, (b) $\NH/\NHIstar$, (c) $\tauHI$, and (d) $\Ts$, respectively, for the region shown in Figure~\ref{fig:Perseus_maps_tauHI_Ts}. We assumed the distance to the Perseus cloud as $300\,\Upc$ \citep{2008hsf1.book..308B}, and the mass is calculated by including hydrogen and helium. We found the mass-weighted mean values as $\langle\NH\rangle\sim1.7\times10^{21}\,\UCND$, $\langle\NH/\NHIstar\rangle\sim1.6$, $\langle\tauHI\rangle\sim0.92$, and $\langle\Ts\rangle\sim97\,\UK$. The ratio of the $\HI$ mass with $\tauHI<1$ is ${\sim}37\%$ of the total mass and as shown in Figure~\ref{fig:Perseus_corr_tau353_WHI} optically thick $\HI$ cannot be ignored. In \citet{2014ApJ...796...59F} we found that $\langle\Ts\rangle\sim30\,\UK$ around the MBM 53, 54, 55/HLCG 92{--}35 clouds, which is somewhat lower than that for the Perseus region. It is possible that the Perseus cloud is located in the Gould's belt, which contains heat sources including some SFRs and OB associations.

\begin{figure*}[]
    \centering
    \includegraphics[scale=1]{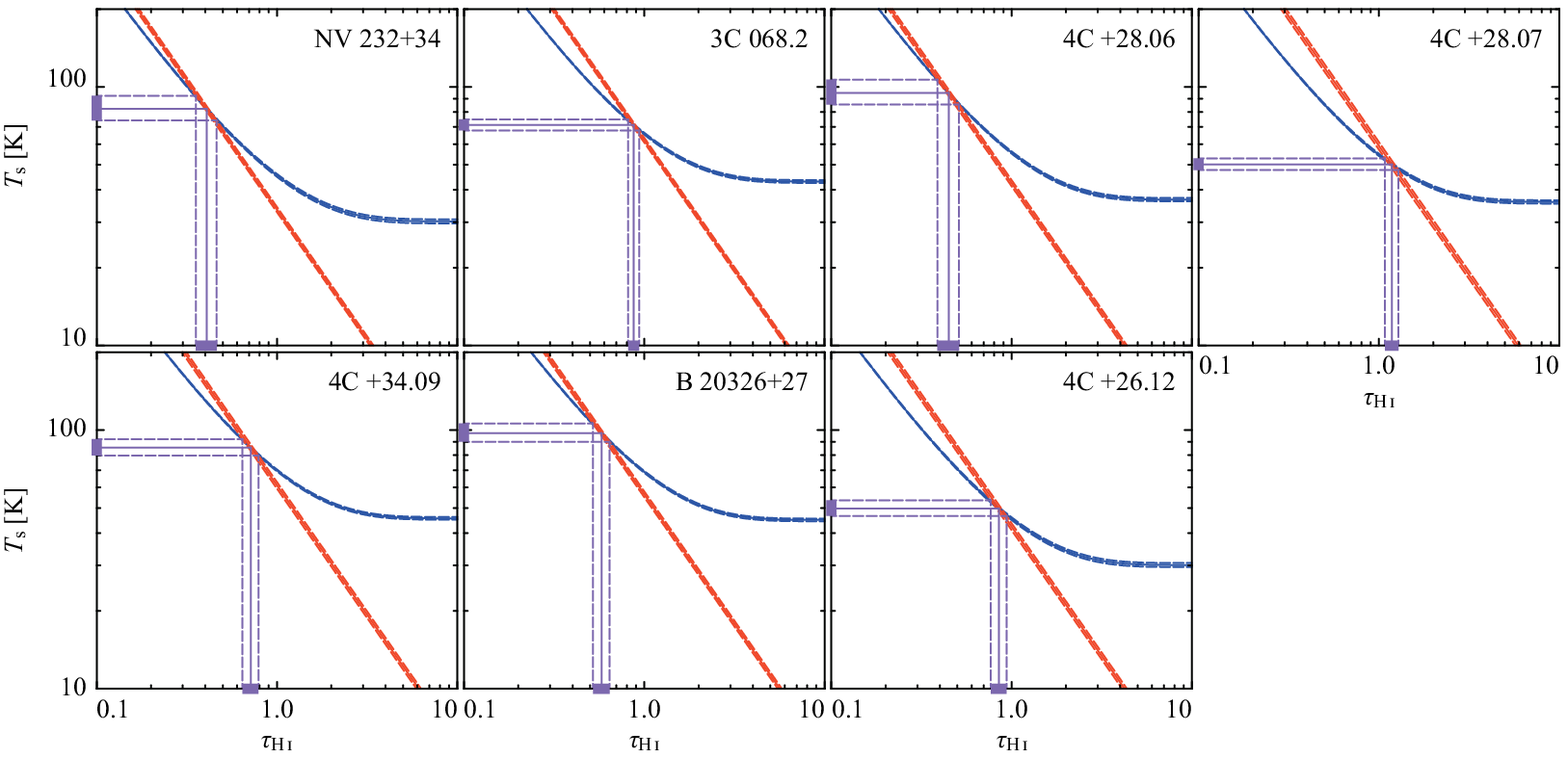}
    \caption{Examples of the $\tauHI$ and $\Ts$ estimations. As examples we show the results toward background radio sources described in \citet{2014ApJ...793..132S}. Blue solid lines are calculated by Equation~(\ref{eq:radiation_transfer}) and red ones by Equation~(\ref{eq:def_tauHI}). The solutions of $\tauHI$ and $\Ts$ (purple solid lines) are obtained in terms of the intersections of the red and blue lines. Dashed lines indicate the uncertainty ranges for each solid line.}
    \label{fig:Perseus_curves_tauHI_Ts}
\end{figure*}

\begin{deluxetable*}{ccccccccc}
\tablewidth{\hsize}
\tablecaption{List of radio continuum sources in Table 1 of \citet{2014ApJ...793..132S}}
\tablenum{2}
\tablehead{\colhead{} & \colhead{} & \colhead{} & \multicolumn{5}{c}{present study} & \multicolumn{1}{c}{\citet{2014ApJ...793..132S}}\\
\colhead{Name} & \multicolumn{2}{c}{Position} & \colhead{$\taud$} & \colhead{$\NH$} & \colhead{$\Ts$} & \colhead{$\tauHI$} & \colhead{$\tauHI\times\dVHI$} & \colhead{$\int\tauHI dV$}\\
\colhead{} & \colhead{$\RA$} & \colhead{$\Dec$} & \colhead{$[10^{-5}]$} & \colhead{$[10^{21}\,\UCND]$} & \colhead{$[\UK]$} & \colhead{} & \colhead{$[\UVel]$} & \colhead{$[\UVel]$}\\
\colhead{(a)} & \colhead{(b)} & \colhead{(c)} & \colhead{(d)} & \colhead{(e)} & \colhead{(f)} & \colhead{(g)} & \colhead{(h)} & \colhead{(i)}}
\startdata
NV 0232+34 & $2^{\text{h}}\,32^{\text{m}}\,28\fs72$ & $+34^{\circ}\,24^{\text{m}}\,06\fs08$ & 0.5 & 0.8 & $82_{-8}^{+10}$ & $0.4{\pm}0.1$ & ${\phn}5.1{\pm}0.7$  & ${\phn}1.89{\pm}0.14$\\
3C 068.2   & $2^{\text{h}}\,34^{\text{m}}\,23\fs87$ & $+31^{\circ}\,34^{\text{m}}\,17\fs62$ & 0.9 & 1.2 & $71_{-3}^{+4}$  & $0.9{\pm}0.1$ & ${\phn}9.0{\pm}0.6$  & ${\phn}4.77{\pm}0.14$\\
4C +28.06  & $2^{\text{h}}\,35^{\text{m}}\,35\fs41$ & $+29^{\circ}\,08^{\text{m}}\,57\fs73$ & 0.7 & 1.0 & $95_{-9}^{+12}$ & $0.4{\pm}0.1$ & ${\phn}5.8{\pm}0.8$  & ${\phn}3.67{\pm}0.04$\\
4C +28.07  & $2^{\text{h}}\,37^{\text{m}}\,52\fs42$ & $+28^{\circ}\,48^{\text{m}}\,09\fs16$ & 1.1 & 1.4 & $50_{-2}^{+3}$  & $1.2{\pm}0.1$ & $15.7_{-1.3}^{+1.4}$ & ${\phn}4.43{\pm}0.15$\\
4C +34.09  & $3^{\text{h}}\,01^{\text{m}}\,42\fs38$ & $+35^{\circ}\,12^{\text{m}}\,20\fs84$ & 1.1 & 1.4 & $85_{-6}^{+7}$  & $0.7{\pm}0.1$ & ${\phn}8.9{\pm}0.9$  & ${\phn}4.05{\pm}0.04$\\
4C +30.04  & $3^{\text{h}}\,11^{\text{m}}\,35\fs19$ & $+30^{\circ}\,43^{\text{m}}\,20\fs62$ & 2.0 & 2.2 & \nodata         & \nodata       & \nodata              & ${\phn}7.62{\pm}0.08$\\
B 20326+27 & $3^{\text{h}}\,29^{\text{m}}\,57\fs69$ & $+27^{\circ}\,56^{\text{m}}\,15\fs64$ & 1.2 & 1.4 & $97_{-7}^{+9}$  & $0.6{\pm}0.1$ & ${\phn}8.1{\pm}0.9$  & ${\phn}4.50{\pm}0.04$\\
3C 092     & $3^{\text{h}}\,40^{\text{m}}\,08\fs55$ & $+32^{\circ}\,09^{\text{m}}\,02\fs32$ & 8.1 & 6.5 & \nodata         & \nodata       & \nodata              & $13.96{\pm}0.99     $\\
3C 093.1   & $3^{\text{h}}\,48^{\text{m}}\,46\fs93$ & $+33^{\circ}\,53^{\text{m}}\,15\fs41$ & 2.2 & 2.4 & \nodata         & \nodata       & \nodata              & ${\phn}8.93{\pm}0.23$\\
4C +26.12  & $3^{\text{h}}\,52^{\text{m}}\,04\fs36$ & $+26^{\circ}\,24^{\text{m}}\,18\fs11$ & 0.8 & 1.1 & $50_{-3}^{+4}$  & $0.9{\pm}0.1$ & $11.7_{-1.1}^{+1.2}$ & ${\phn}4.87{\pm}0.04$\\
\enddata
%\tablenotetext{}{}
\tablecomments{
    (a), (b), and (c): Names and coordinates of the radio continuum sources listed in Table 1 of \citet{2014ApJ...793..132S}.\\
    (d): $\taud$ values toward the sources.\\
    (e): $\NH$ values toward the sources calculated by Equation~(\ref{eq:calc_NH}).\\
    (f) and (g): $\HI$ spin temperature and $\HI$ optical depth calculated by Equations (\ref{eq:radiation_transfer}) and (\ref{eq:def_tauHI}).\\
    (h): Velocity-integrated $\HI$ optical depth derived from column (g) and $\dVHI$.\\
    (i): Total Velocity-integrated $\HI$ optical depth derived from the Gaussian parameters of $\tauHI$ spectra of  \citet{2014ApJ...793..132S}.\label{tab:list_Stanimirovic+2014}
}
\end{deluxetable*}

Figure~\ref{fig:Perseus_curves_tauHI_Ts} shows examples of the $\tauHI$ and $\Ts$ solutions. The curves of Equation~(\ref{eq:radiation_transfer}) (blue) and Equation~(\ref{eq:def_tauHI}) (red), and the solutions of $\tauHI$ and $\Ts$ (purple) are drawn on the $\tauHI\text{--}\Ts$ plane. We show the solutions toward the background radio sources described in \citet{2014ApJ...793..132S}. As described above, the $\HI$ spectra toward the directions are spatially interpolated. The blue and red dashed curves indicate $1\,\sigma$ uncertainties for each curve, and we defined the uncertainties in the solutions (purple lines) as the intersections of the dashed curves. Table~\ref{tab:list_Stanimirovic+2014} is the results of the calculations. For each data point the solutions of $\tauHI$ and $\Ts$ are well determined.

\subsection{$\XCO$ estimation}\label{sec:XCO_estimation}

The $\XCO$ factor, which is an empirical conversion factor, has been used in order to estimate $\Htwo$ column number density ($\NHtwo$) from the velocity-integrated intensity of the $\CO$ line ($\WCO$). Conventionally, $\XCO\sim(1{\text{--}}2)\times10^{20}\,\UXCO$ is assumed as an typical value for the Galaxy \citep[e.g.,][]{2013ARA&A..51..207B}. Here, since $\NH$ is obtained with a higher accuracy, we are able to determine $\XCO$ by using the {\Planck} data and $\WCO$ data.

\begin{figure*}[]
    \centering
    \includegraphics[scale=1]{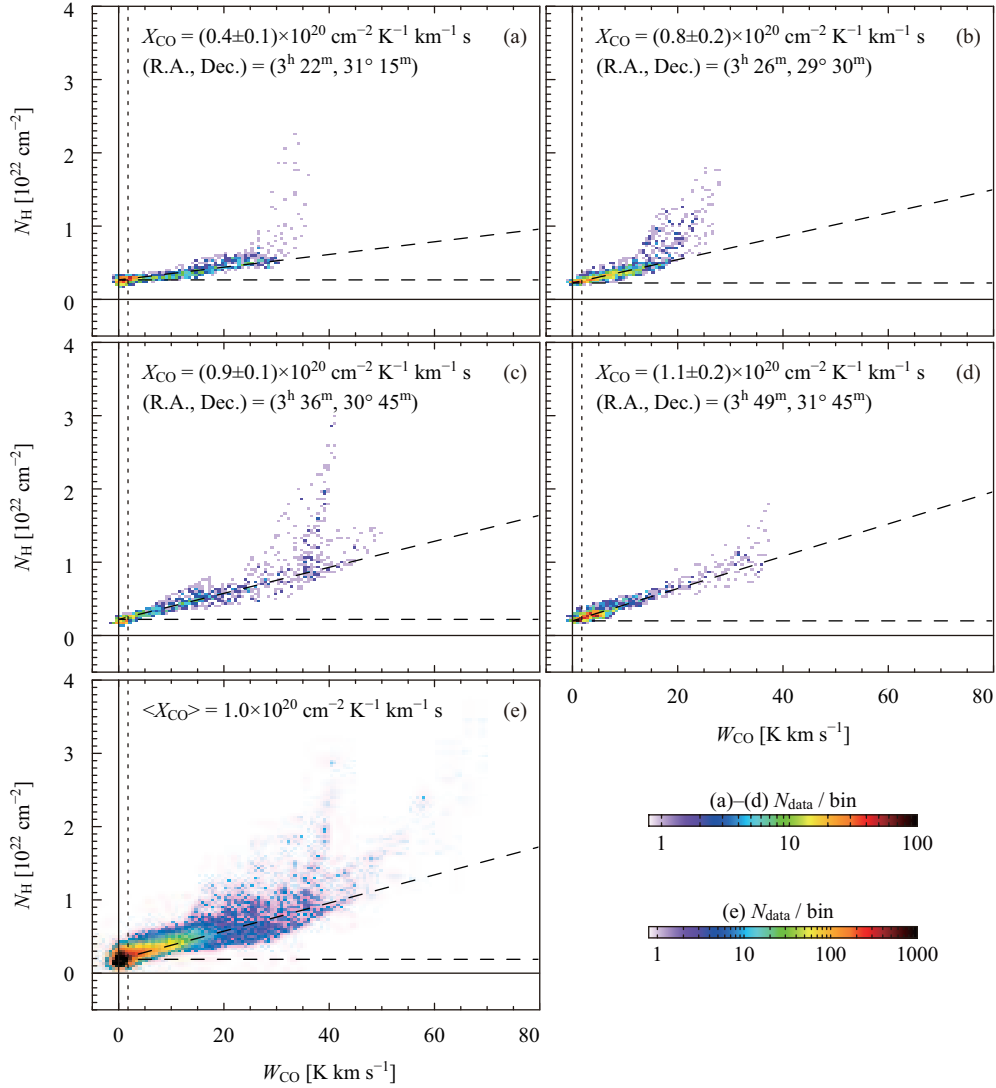}
    \caption{(a)--(d) Examples of $\WCO${--}$\NH$ correlation plots (density plots) using the data points inside the $r=1\,\Udeg$ windows (A){--}(D) shown in Figure~\ref{fig:Perseus_map_XCO}. (e) The same as panels (a)--(d) but using the all data points shown in Figure~\ref{fig:Perseus_map_XCO}(a) and (b). The outlier-robust linear regression method is used to fit the data in order to avoid effects due to saturated points. The data points of $\WCO\ge3\,\sigma$ are used. The dotted lines indicate $\WCO=3\,\sigma$. The dashed lines show the resulting slopes and intercepts of the regressions, and $\XCO$ is estimated by Equation~(\ref{eq:corr_WCO_NH}) by using these slopes. The average value of $\XCO$ is calculated as $\langle\XCO\rangle\sim1.0\times10^{20}\,\UXCO$ from panel (e).}
    \label{fig:Perseus_corr_WCO_NH}
\end{figure*}

\begin{figure*}[]
    \centering
    \includegraphics[scale=1]{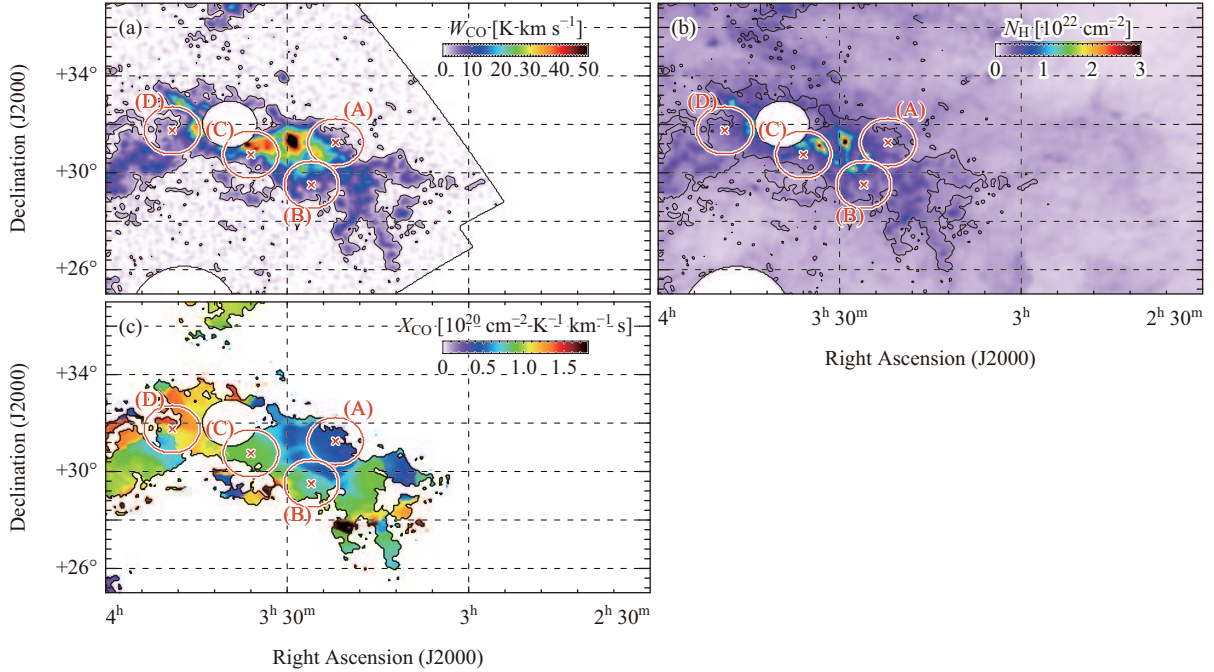}
    \caption{(a) The map of $\WCO$ overlaid with $r=1^{\circ}$ windows (A){--}(D). (b) The map of $\NH$ overlaid with the windows (A){--}(D). (c) A map of the estimated spatial distribution of $\XCO$ factor. In Figure~\ref{fig:Perseus_corr_WCO_NH}(a)--(d) we show examples of $\WCO$\text{--}$\NH$ correlation plots using the data points inside the windows (A){--}(D) used to compute $\XCO$ at their center positions (X-points). The average value of $\XCO$ is $\langle\XCO\rangle=1.0\times10^{20}\,\UXCO$ (Figure~\ref{fig:Perseus_corr_WCO_NH}(e)).}
    \label{fig:Perseus_map_XCO}
\end{figure*}

Figure~\ref{fig:Perseus_corr_WCO_NH} indicates correlation plots between $\WCO$ ($x$--axis) and $\NH$ ($y$--axis). Fitting the distribution of the data points and calculating the slope and the intercept, we can separate the atomic component and the molecular component from $\NH$.
\begin{eqnarray}
    \NH&=\NHI+2\,\NHtwo\nonumber\\
    \NHtwo&=\XCO\times\WCO\nonumber
\end{eqnarray}
So, we obtain the following,
\begin{eqnarray}
    \NH&=(2\,\XCO)\,\WCO+\NHI\nonumber\\
    &\equiv(\text{slope})\times\WCO+(\text{intercept}).\label{eq:corr_WCO_NH}
\end{eqnarray}
The intercept gives the atomic component, and the slope gives the molecular component and the $\XCO$ factor ($=\text{slope}/2$). Note that the contribution of the atomic component is a mean value, and we assumed that it is constant against $\WCO$. We used an outlier-robust linear regression method in order to avoid the effect due to the data points where $\WCO$ is saturated against $\NH$. For this purpose, we utilized the ROBUST\_LINEFIT routine provided in The IDL Astronomy User's Library \citep{1993ASPC...52..246L}. In the present study, we estimated the spatial distribution of the $\XCO$ factor by the procedure described below.
\begin{enumerate}
    \item We prepare a ``window'', which extracts the data points within a $1$-degree radius from a certain point. This radius is determined in order to ensure a sufficient number of data points to be fitted.
    \item By using the data points inside the window, we estimate the $\XCO$ factor from the $\WCO${--}$\NH$ plot .
    \item We regard the resulting $\XCO$ as the value of the center point of the window.
    \item The $\XCO$ map can be obtained by iteratively calculating the $\XCO$ factor for all the data points while moving the center of the window pixel by pixel (this process is similar to image convolution).
\end{enumerate}
Figure~\ref{fig:Perseus_map_XCO}(a) and (b) shows the map of $\WCO$ and $\NH$ (already shown in Figure~\ref{fig:Perseus_maps_total}(a) and Figure~\ref{fig:Perseus_maps_NHratio}(a)), but only Mask-(d) and (e) are applied in order to compare with previous studies \citep[][described below]{2012ApJ...748...75L,2014ApJ...784...80L}. The four overlaid circles (A)--(D) indicate the examples of the ``window'', corresponding to the panels (a)--(d) of Figure~\ref{fig:Perseus_corr_WCO_NH}, respectively. The resulting $\XCO$ map is shown in Figure~\ref{fig:Perseus_map_XCO}(c). We found the $\XCO$ factor significantly varies within the Perseus cloud, $\XCO\sim(0.3\text{--}2.0)\times10^{20}\,\UXCO$ with a typical uncertainty of $10\text{--}50\%$. The $\XCO$ variation within molecular clouds is discussed in some previous studies \citep[e.g.,][]{1998ApJ...504..290M,2013MNRAS.436.1152C,2014ApJ...784...80L}, and the present result agrees with these results. We also found the average value is $\langle\XCO\rangle\sim1.0\times10^{20}\,\UXCO$ (Figure~\ref{fig:Perseus_corr_WCO_NH}(e)), which is consistent with the empirical and typical value of the Galaxy, $(1{\text{--}}2)\times10^{20}\,\UXCO$. On the other hand, \citet{2012ApJ...748...75L,2014ApJ...784...80L} also estimated a spatial distribution the $\XCO$ factor in the Perseus molecular cloud. However, they obtained the average value of $\langle\XCO\rangle\sim0.3\times10^{20}\,\UXCO$, which is ${\sim}1/3$ of our result. We will explain this discrepancy between these two results on the $\XCO$ factor in the next section.

\section{Discussions}\label{sec:Discussions}

\subsection{$\XCO$}\label{subsec:XCO}

As mentioned above, \citet{2012ApJ...748...75L,2014ApJ...784...80L} also estimated the spatial distribution of the $\XCO$ factor in the Perseus molecular cloud. Their procedure for the $\XCO$ estimation is briefly described below:
\begin{enumerate}
    \item $\NH$ is estimated by using $\AV$ as a tracer of the total amount of hydrogen. They used the dust-to-gas ratio (DGR) as the conversion factor,
          \begin{eqnarray}
              \NH=\frac{\AV}{\text{DGR}},\ \ \ \ \ \text{DGR}=1.1\times10^{-21}\,\Umag^{-1}\,\Ucm^{2}.\nonumber
          \end{eqnarray}
    \item $\NHtwo$ is calculated by subtracting $\NHI$ from $\NH$,
          \begin{eqnarray}
              \NHtwo=\frac{1}{2}(\NH-\NHI).\label{eq:Lee_calc_NH2}
          \end{eqnarray}
          Note that in \citet{2012ApJ...748...75L,2014ApJ...784...80L} $\tauHI=0$ is assumed, that is, $\NHI=\NHIstar$ is assumed.
    \item The resulting $\XCO$ map is obtained by dividing $\NHtwo$ by $\WCO$ for each data point.
\end{enumerate}
The available $\AV$ map (``COMPLETE survey'' \citep{2001AaA...377.1023L} based on the 2MASS data), however, covers only the central region of the cloud, hence they made a wide-area ``simulated'' $\AV$ map by using {\IRAS} data:
\begin{enumerate}
    \item The wide-area dust temperature map was estimated from the ratio of the intensities of the IRIS $60, 100\,\Umicron$ maps. Effect from Very Small Grains (VSGs) was considered for the IRIS $60\,\Umicron$ map.
    \item The wide-area map of optical depth at $100\,\Umicron$ ($\tau_{100}$) was derived from the ratio of the IRIS $100\,\Umicron$ map to the intensity at $100\,\Umicron$ of the blackbody radiation at derived dust temperature. The zero point offset in $\tau_{100}$ was also corrected.
    \item The wide-area ``simulated'' $\AV$ map was obtained from the $\tau_{100}$ map by using the conversion factor obtained by the correlation between $\tau_{100}$ and the COMPLETE $\AV$ map.
\end{enumerate}

\begin{figure*}[]
    \centering
    \includegraphics[scale=1]{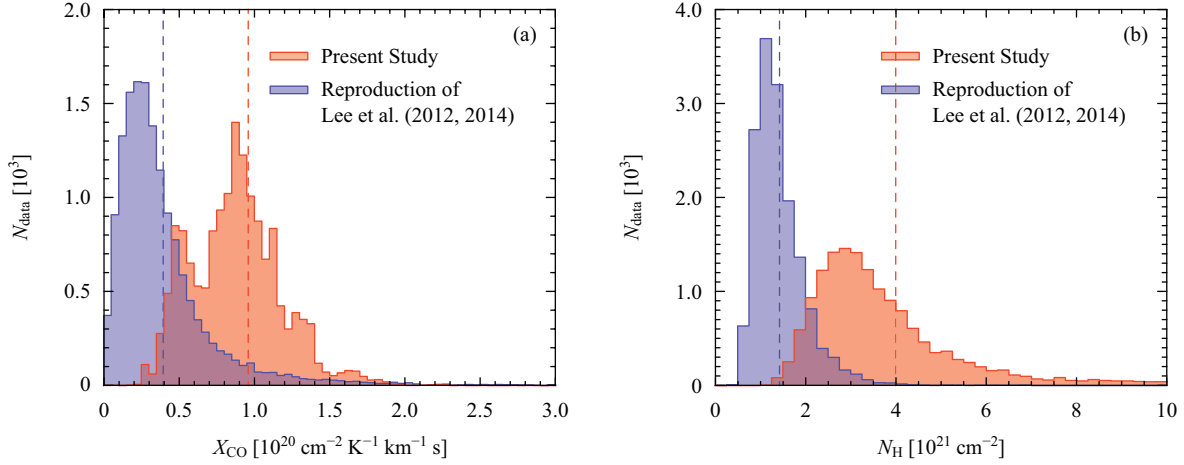}
    \caption{Histograms of (a) $\XCO$ and (b) $\NH$ for the region shown in Figure~(\ref{fig:Perseus_map_XCO}). The orange ones are obtaind in the present study and purple ones are obtained by replicating the procedures described in \citet{2012ApJ...748...75L,2014ApJ...784...80L}. The dashed lines denote the average values of each histogram.}
    \label{fig:Perseus_histograms_XCO_NH}
\end{figure*}

Figure~\ref{fig:Perseus_histograms_XCO_NH}(a) shows histograms of $\XCO$ derived in Section~\ref{sec:XCO_estimation} and $\XCO$ reproduced by the same procedure as \citet{2012ApJ...748...75L,2014ApJ...784...80L}. Note that the data points used are the same as Figure~\ref{fig:Perseus_map_XCO}, not the same as \citet{2012ApJ...748...75L,2014ApJ...784...80L}. The average values are $\langle\XCO\rangle\sim1.0\times10^{20}\,\UXCO$ and $\langle\XCO\rangle_{\text{Lee}}\sim 4.0\times10^{19}\,\UXCO$ for the present study and the Lee's result, respectively, and the result of \citet{2012ApJ...748...75L,2014ApJ...784...80L} is well reproduced. In Figure~\ref{fig:Perseus_histograms_XCO_NH}(b) we show histograms of $\NH$ derived by using Equation~(\ref{eq:calc_NH}) and the same method as \citet{2012ApJ...748...75L}. We found $\langle\NH\rangle\sim4.3\times10^{21}\,\UCND$ and $\langle\NH\rangle_{\text{Lee}}\sim1.4\times10^{21}\,\UCND$, hence \citet{2012ApJ...748...75L} underestimated $\NH$ to be ${\sim}40\%$.

\begin{figure*}[]
    \centering
    \includegraphics[scale=1]{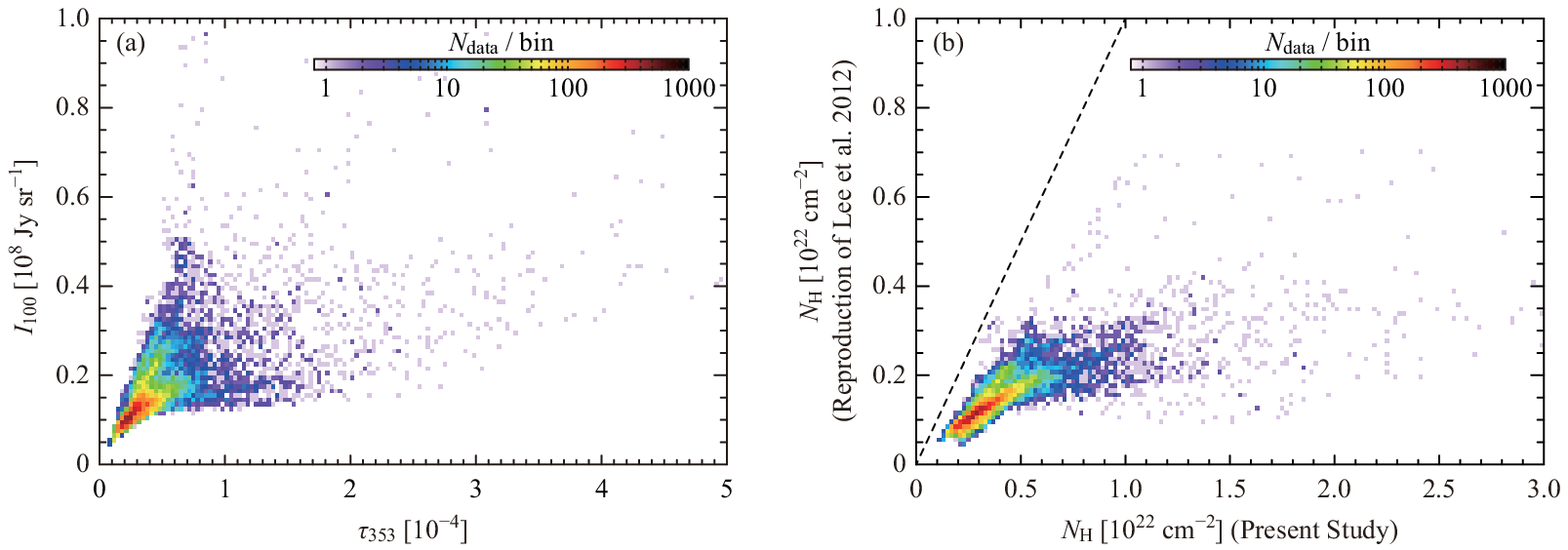}
    \caption{
        Correlation plots (density plots) between (a) $\taud$ and $\Ihundred$, (b) $\NH$ calculated by Equation~(\ref{eq:calc_NH}) and $\NH$ calculated by the same method as \citet{2012ApJ...748...75L}. The dashed line in panel (b) indicates the $1{:}1$ relationship.
    }
    \label{fig:Perseus_corr_comparing_with_Lee+}
\end{figure*}

Here, we will quantitatively examine the difference in $\XCO$ between the present study and \citet{2012ApJ...748...75L,2014ApJ...784...80L}. Since $\XCO=\NHtwo/\WCO$, the underestimation of $\XCO$ in \citet{2012ApJ...748...75L,2014ApJ...784...80L} is caused by the underestimation of $\NHtwo$ in Equation~(\ref{eq:Lee_calc_NH2}). Therefore, we investigated the cause of the underestimation of $\NH$ and $\NHI$ in Equation~(\ref{eq:Lee_calc_NH2}). Figure~\ref{fig:Perseus_corr_comparing_with_Lee+}(a) is a correlation plot between $\taud$ and $\Ihundred$. The correlation coefficient is ${\sim}0.5$ and we found that $\Ihundred$ is not a good tracer of $\NH$. Figure~\ref{fig:Perseus_corr_comparing_with_Lee+}(b) shows a correlation between $\NH$ calculated by Equation~(\ref{eq:calc_NH}) and $\NH$ calculated by the same method as \citet{2012ApJ...748...75L}. The dashed line indicates the $1{:}1$ relationship. The latter is clearly underestimated (${\sim}35\%$ on average) against the former. Separately from that, in \citet{2012ApJ...748...75L,2014ApJ...784...80L} the effect of $\HI$ optical depth was not considered. As shown in Section~\ref{Hydrogen_Amount_Estimation} the effect of $\tauHI$ is $\NHI/\NHIstar\sim1.6$. Therefore, $\NHI$ is underestimated to be (at least) ${\sim}62\%$ on average. In addition, the integrated velocity range of $\HI$ spectra is from $-5\,\UVel$ to $+15\,\UVel$ in \citet{2012ApJ...748...75L}. This corresponds to ${\sim}76\%$ of the total integrated intensity of the mean spectrum shown in Figure~\ref{fig:Perseus_PV_meanspectrum}(b). From these, $\NHI$ in Equation~(\ref{eq:Lee_calc_NH2}) is underestimated at least $62\%\times76\%\sim50\%$ or less on average. In \citet{2012ApJ...748...75L,2014ApJ...784...80L}, the right-hand side of Equation~(\ref{eq:Lee_calc_NH2}) is underestimated to be ${\sim}40\%$, and hence, $\XCO$ is underestimated to be ${\sim}40\%$ against our result.

\begin{figure}[p]
    \centering
    \includegraphics[scale=1]{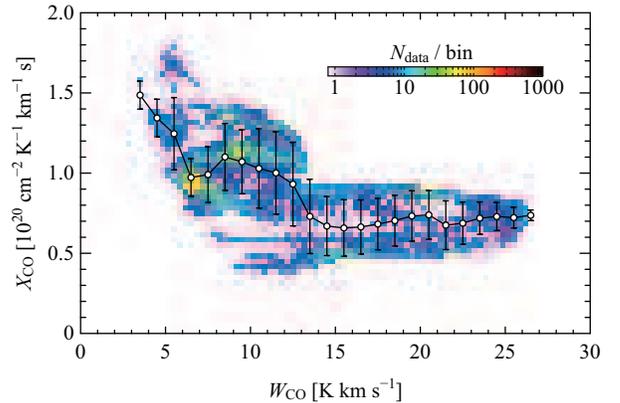}
    \caption{
        A correlation plot (density plot) between $\WCO$ and $\XCO$. In order to compare with $\XCO$, we spatially smoothed the $\WCO$ map to a $1\,\Udeg$ effective HPBW. The white circles and the vertical bars indicate the average and the standard deviation of $\XCO$ in each $1\,\UII$ bin.
    }
    \label{fig:Perseus_corr_WCO_XCO}
\end{figure}

Figure~\ref{fig:Perseus_corr_WCO_XCO} shows a correlation between $\WCO$ and $\XCO$ maps. We spatially smoothed the $\WCO$ map to a $1\,\Udeg$ effective HPBW in order to compare with $\XCO$, and we plotted the data points where $\XCO\ge3\,\sigma$. We also show the average and the standard deviation of $\XCO$ in each $1\,\UII$ bin as the white circles and the vertical bars. From this plot, we found an anti-correlation relationship between the two variables. This anti-correlation reflects that in the diffuse (low-$\WCO$) region $\CO$ molecule is photo-dissociated more effectively than $\Htwo$, and therefore $\XCO$ increases. This trend is consistent with the result in \citet{2013MNRAS.436.1152C,2014A&A...566A.120S}, and we can say that the spatial distribution of $\XCO$ can be well calculated by using the present method.

\subsection{Mass}\label{subsec:mass}

The total mass of the hydrogen gas including $\HI$, $\Htwo$, and helium is derived as $\MH=1.8\times10^{5}\,\Msol$ by using $\NH$ calculated by Equation~(\ref{eq:calc_NH}). We assumed the distance to the Perseus cloud as $300\,\Upc$ \citep{2008hsf1.book..308B}. On the other hand, the mass of the molecular hydrogen gas is calculated as $\MHtwo=2.5\times10^{4}\,\Msol$ by using $\WCO$ and the $\XCO$ map obtained in Section~\ref{sec:XCO_estimation}. From these masses, we calculate the mass of the atomic hydrogen gas as $\MHI=\MH-\MHtwo=1.5\times10^{5}\,\Msol$. This indicates that the molecular cloud is surrounded by the atomic gas whose mass is by an order of magnitude larger than that of the molecular one, and it also indicates that the atomic gas is the most principal component of the ISM. This is consistent with the result for the MBM 53, 54, 55/HLCG 92{--}35 region described in \citet{2014ApJ...796...59F}. Note that the virial mass of the $\HI$ gas around the Perseus cloud is calculated as ${\sim}2\times10^{6}\Msol$ if we assume that the gas around the Perseus cloud is a sphere with a radius of $50\,\Upc$ and a velocity width of $15\,\UVel$. This is an order of magnitude larger than the estimated mass above ($1.5\times10^{5}\,\Msol$), and hence, it is not gravitationally bound. The virial mass of the $\CO$ cloud is ${\sim}1\times10^{5}\,\Msol$ if we assume that its radius is $25\,\Upc$ and its velocity width is $5\,\UVel$. The molecular mass estimated above ($2.5\times10^{4}\,\Msol$) is ${\sim}1/4$ of this virial mass, and the cloud traced by $^{12}\CO(J{=}1{\text{--}}0)$ line is also not gravitationally bound. Since the molecular cloud is denser than the atomic cloud, the mass of the molecular cloud is relatively closer to its virial mass than that of the atomic cloud.

\subsection{$\tauHI$}\label{subsec:tauHI}

\begin{figure}[]
    \centering
    \includegraphics[scale=1]{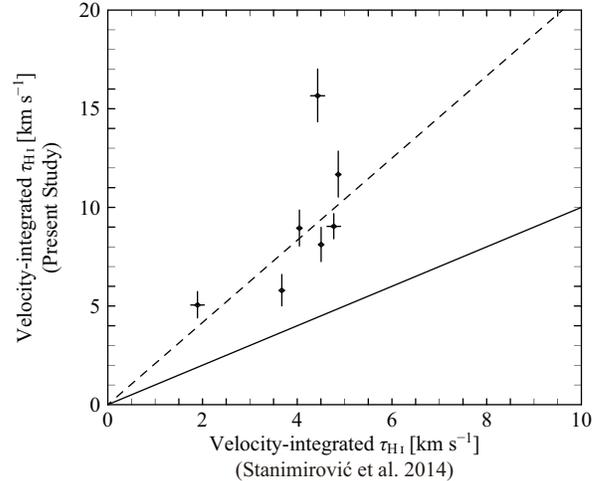}
    \caption{A correlation plot between velocity-integrated values of $\tauHI$ derived in \citet{2014ApJ...793..132S} ($x$-axis) and $\tauHI\times\dVHI$ derived in the present study ($y$-axis). The solid line indicates the 1{:}1 relationship between both variables. The dashed line is the result of a linear regression through the origin. Its slope is $2.1$.}
    \label{fig:Perseus_Stanimirovic+2014_corr}
\end{figure}

As one of the previous studies on the atomic gas in the Perseus region, \citet{2014ApJ...793..132S} calculated the $\HI$ optical depth toward 26 extra-Galactic radio continuum sources (such as quasars). They calculated $\tauHI$ as a function of velocity toward each radio source by using the method described in \citet{2003ApJS..145..329H}. The calculated $\tauHI$ spectra were fitted with a sum of Gaussian functions, and the peak $\tauHI$ values and the Gaussian FWHMs (full width at half maximum) of each velocity component were derived. However, they found the optically-thick $\HI$ gas only toward ${\sim}15\%$ of lines of sight. We test the results in \citet{2014ApJ...793..132S} and \citet{2014ApJ...796...59F,2015ApJ...798....6F}, which differ a factor of several. To do this, we compare $\tauHI$ derived in \citet{2014ApJ...793..132S} and that obtained in Section~\ref{sec:tauHI_Ts_Estimation}. Since our $\tauHI$ corresponds to the average value within given velocity width ($\dVHI$) (see Section~\ref{Hydrogen_Amount_Estimation}) we compare the following two values toward each radio source located in the region we analyzed; (1) the sum of the areas of each Gaussian component of the $\tauHI$ profiles calculated in \citet{2014ApJ...793..132S}, and (2) the products of our $\tauHI$ and $\dVHI$. The results are listed in the column 8 and 9 of Table~\ref{tab:list_Stanimirovic+2014}. Note that 4C +30.04, 3C 092, and 3C 093.1 are located at the masked area, therefore $\tauHI$ cannot be calculated by our method toward them. Although there is a rough positive correlation between them, they differ $2{\text{--}}4$ times as a concrete numerical value. The correlation between them is plotted in Figure~\ref{fig:Perseus_Stanimirovic+2014_corr}. The solid line indicates the one-to-one relationship ($\text{slope}=1$), and the dashed line is the best-fit regression line through the origin ($\text{slope}=2.1$). The results in \citet{2014ApJ...793..132S} are systematically smaller than our results, and it is obvious that there is a discrepancy between these two results. This discrepancy can, however, be explained by characteristics of the data used to derive $\tauHI$ in the present study and \citet{2014ApJ...793..132S}.

In \citet{2014ApJ...793..132S} $\tauHI$ is calculated based on the $\HIcm$ absorption spectrum toward the extra-Galactic point sources. Since these point sources have infinitesimal apparent size, the absorption by the $\HI$ gas actually occurs toward the very direction of each source. However, \citet{2012ApJ...759...35I} and \citet{2006ApJ...652.1339M} revealed that the cold neutral medium (CNM), which is an optically-thick component of the $\HI$ gas, has highly filamentary and spatially inhomogeneous structure. Based on the three-dimensional numerical simulation of magnetohydrodynamics (MHD) \citep{2012ApJ...759...35I}, we found that the CNM accounts for typically ${\sim}30\%$ of the 2D-projected area \citep{2017arXiv170107129F}. Hence, the probability with which the CNM affects the absorption toward the point sources is small. Therefore, unless the observational beam size is not infinitesimal, absorption measurements tend to detect only warm neutral medium (WNM) and tend to regard the optical depth of the WNM as the average value within the beam. On the other hand, we use the emission data of gas and dust, and they include information from both the CNM and the WNM within the large beam (${\gtrsim}4\,\Uarcmin$). Therefore, we can say that the discrepancy of $\tauHI$ is due to the difference between the measurement methods (absorption toward point sources or emission). We independently confirm this result by using the numerical simulation (``synthetic'' observations) separately in \citet{2017arXiv170107129F}.

\section{Conclusions}\label{sec:Conclusions}

In the present paper, we discussed the amount of the interstellar hydrogen gas and $\HI$ optical depth in the Perseus region by using the dust emission parameters obtained {\Planck}, $\HI$ $\HIcm$ line, and $\CO$ line data. The results are as follows:
\begin{enumerate}
    \item The distributions of the data points on the $\taud${--}$\WHI$ plot systematically vary by the change in $\Td$. This is consistent with the results in \citet{2014ApJ...796...59F,2015ApJ...798....6F}, which is that $\HI$ optical depth cannot be ignored for the low-$\Td$ points. However, since the distance to the Perseus region is larger than that to the region analyzed in \citet{2014ApJ...796...59F,2015ApJ...798....6F} and there exist the SFRs, the $\taud${--}$\WHI$ plot is somewhat complicated as compared with the local clouds. As described in Section~\ref{Hydrogen_Amount_Estimation}, the previous studies \citep{2012ApJ...746...82F,2015ApJ...798....6F} showed that the total hydrogen amount can be explained without assuming the presence of ``$\CO$-dark $\Htwo$ gas''. Therefore, we consider that the bad correlation between $\taud$ and $\WHI$ for the low-$\Td$ points is due to optically-thick $\HI$ gas.
    \item In order to consider the dust evolution at high density regions, we calibrated the $\taud${--}$\NH$ relationship by using the extinction at $J$-band. As with \citet{2013ApJ...763...55R}, we confirmed that $\taud$ becomes larger as a function of the $1.3$-th power of $\NH$ ($\alpha=1.3$). In addition, we reconsidered the reference values of $\taud$ and $\NH$ in Equation~(\ref{eq:Fukui+2015_eq7}) by referring to the results in \citet{2014ApJ...796...59F}. The calibrated $\taud${--}$\NH$ relationship (Equation~(\ref{eq:calc_NH})) yields ${\sim}20\%$ smaller $\NH$ than that in the case of $\alpha=1.0$. By using this relationship we can calculate $\NH$ from $\taud$ more accurately, taking the dust evolution into account.
    \item Compared with the present method, the conventional method which assumes that the $\HI$ gas is optically-thin, underestimates the amount of the hydrogen gas to be ${\sim}62\%$ in the Perseus region. Optical depth of the $\HI$ gas ($\tauHI$) and spin temperature ($\Ts$) can be calculated, and we obtain $\langle\tauHI\rangle\sim0.92$. These results support that there exists a large amount of optically-thick $\HI$ gas around the molecular clouds. The arguments by \citet{2014ApJ...796...59F,2015ApJ...798....6F} still hold in the region where the density is relatively high and there exist the SFRs.
    \item By using $\NH$ calculated from $\taud$ and the $\CO$ intensity ($\WCO$), we estimated the spatial distribution of $\XCO$. We obtained $\langle\XCO\rangle\sim1.0\times10^{20}\,\UXCO$, which is consistent with the conventional value in the Galaxy, $\XCO\sim(1{\text{--}}2)\times10^{20}\,\UXCO$. The relative uncertainty in $\XCO$ is smaller than conventional estimation and that in previous studies, typically $10\text{--}50\%$. Although the result in \citet{2012ApJ...748...75L,2014ApJ...784...80L} is underestimated to be ${\sim}40\%$ compared to this result, this discrepancy in the $\XCO$ maps can be quantitatively explained by the difference of the data used in order to calculate $\NH$, the effect of the optical depth of $\HI$, and the difference of the integrated velocity ranges of $\HI$ $\HIcm$ spectra.
    \item We compared our $\tauHI$ with that obtained in \citet{2014ApJ...793..132S}. When the optical depth is calculated based on the absorption measurements toward the background point sources, it can be underestimated to be ${\sim}40\%$ on the average compared to that obtained based on the gas/dust emission data. It is revealed that the cold $\HI$ gas (CNM) has highly filamentary distribution by numerical simulation studies \citep[e.g.,][]{2012ApJ...759...35I}. Therefore, the probability with which the optically-thick $\HI$ filaments lie on the infinitesimal apparent-size sources is small, ${\sim}30\%$. The underestimation of $\tauHI$ in absorption measurements is consistent with this picture.
\end{enumerate}

\acknowledgements

This work was financially supported by Grants-in-Aid for Scientific Research (KAKENHI) of the Japan society for the Promotion of Science (JSPS) Grant Numbers 15H05694, 25287035.

This research has made use of NASA's Astrophysics Data System.%http://ads.nao.ac.jp/
This publication utilizes data from Galactic ALFA $\HI$ (GALFA $\HI$) survey data set obtained with the Arecibo $L$-band Feed Array (ALFA) on the Arecibo 305m telescope. Arecibo Observatory is part of the National Astronomy and Ionosphere Center, which is operated by Cornell University under Cooperative Agreement with the U.S. National Science Foundation. The GALFA $\HI$ surveys are funded by the NSF through grants to Columbia University, the University of Wisconsin, and the University of California.%https://purcell.ssl.berkeley.edu/
Based on observations obtained with {\Planck} (\texttt{http://www.esa.int/Planck}), an ESA science mission with instruments and contributions directly funded by ESA Member States, NASA, and Canada.%http://www.cosmos.esa.int/web/planck/planck-data-use
We acknowledge the use of the Legacy Archive for Microwave Background Data Analysis (LAMBDA), part of the High Energy Astrophysics Science Archive Center (HEASARC). HEASARC/LAMBDA is a service of the Astrophysics Science Division at the NASA Goddard Space Flight Center.%https://lambda.gsfc.nasa.gov/
This research has made use of the NASA/IPAC Infrared Science Archive, which is operated by the Jet Propulsion Laboratory, California Institute of Technology, under contract with the National Aeronautics and Space Administration.%http://irsa.ipac.caltech.edu/ack.html
The $\CO$ line data was obtained with the CfA (Harvard-Smithsonian Center for Astrophysics) $1.2\,\Um$ telescope \citep{2001ApJ...547..792D}.
The $\Halpha$ emission data was assembled from the Wisconsin H-Alpha Mapper (WHAM), the Virginia Tech Spectral-Line Survey (VTSS), and the Southern H-Alpha Sky Survey Atlas (SHASSA) \citep{2003ApJS..146..407F}.
%We also used the data sets of  $\Halpha$ emission \citep{2003ApJS..146..407F}, 
Some of the results in this paper have been derived using the HEALPix \citep{2005ApJ...622..759G} package.%http://healpix.sourceforge.net/downloads.php

\appendix

\section{Determination of $\taudref$ and $\NHref$}\label{subsec:determination_of_tau353ref_NHref}

$\taudref$ and $\NHref$ are the normalization constants conveniently introduced in order to nondimensionalize the both sides of Equation~(\ref{eq:Fukui+2015_eq7}). Although in \citet{2015ApJ...798....6F} $\taudref$ and $\NHref$ are determined as $4.77\times10^{-6}$ and $1\times10^{21}\,\UCND$, respectively, we derive our $\taudref$ and $\NHref$ for the Perseus region by using Equation~(\ref{eq:curve_tau353_WHI_theoretical}). By giving $\alpha$, $\tauHI$, and using $\NHref=(1.15\times10^{8})\times\XHI\times\taudref$ \citep{2015ApJ...798....6F} we make the free parameter in Equation~(\ref{eq:curve_tau353_WHI_theoretical}) $\taudref$ alone, and determine it from the $\taud$-$\WHI$ plot. First, we describe the estimation of $\tauHI$ for high-$\Td$ points in the Perseus region.

\begin{figure}[]
    \centering
    \includegraphics[scale=1]{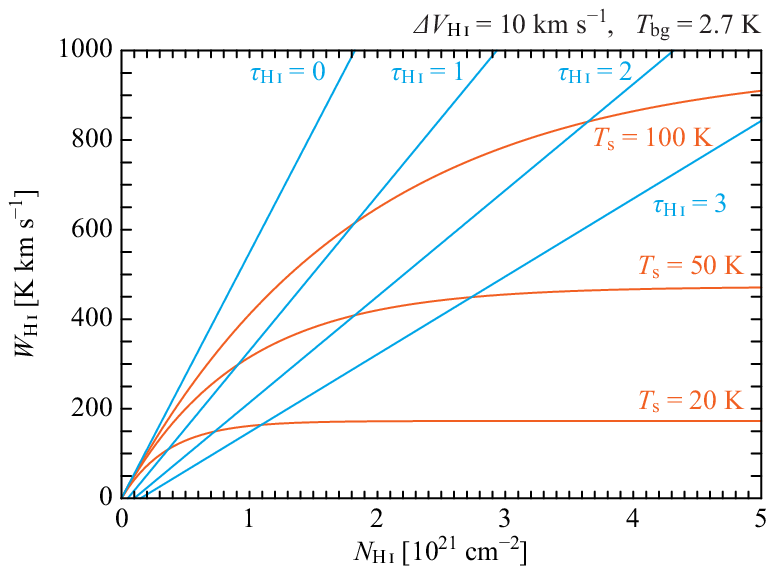}
    \caption{Theoretical curves of $\WHI$ as functions of $\NHI$. The light-blue ones are for the case of fixing $\tauHI=0, 1, 2, 3$, and the orange ones $\Ts=20, 50, 100\,\UK$. $\dVHI=10\,\UVel$ and $\Tbg=2.7\,\UK$ are assumed.}
    \label{fig:Perseus_theoreticalcurves_NHI_WHI}
\end{figure}

Figure~\ref{fig:Perseus_theoreticalcurves_NHI_WHI} shows theoretical relationships between $\NHI$ and $\WHI$ in the cases of fixing $\tauHI=0, 1, 2, 3$ or $\Ts=20, 50, 100\,\UK$. We assumed that $\dVHI=10\,\UVel$ and $\Tbg=2.7\,\UK$. One can see that the variation width in $\WHI$ becomes larger when $\tauHI$ becomes higher even if $\Ts$ is a constant (orange curves).

\begin{figure*}[]
    \centering
    \includegraphics[scale=1]{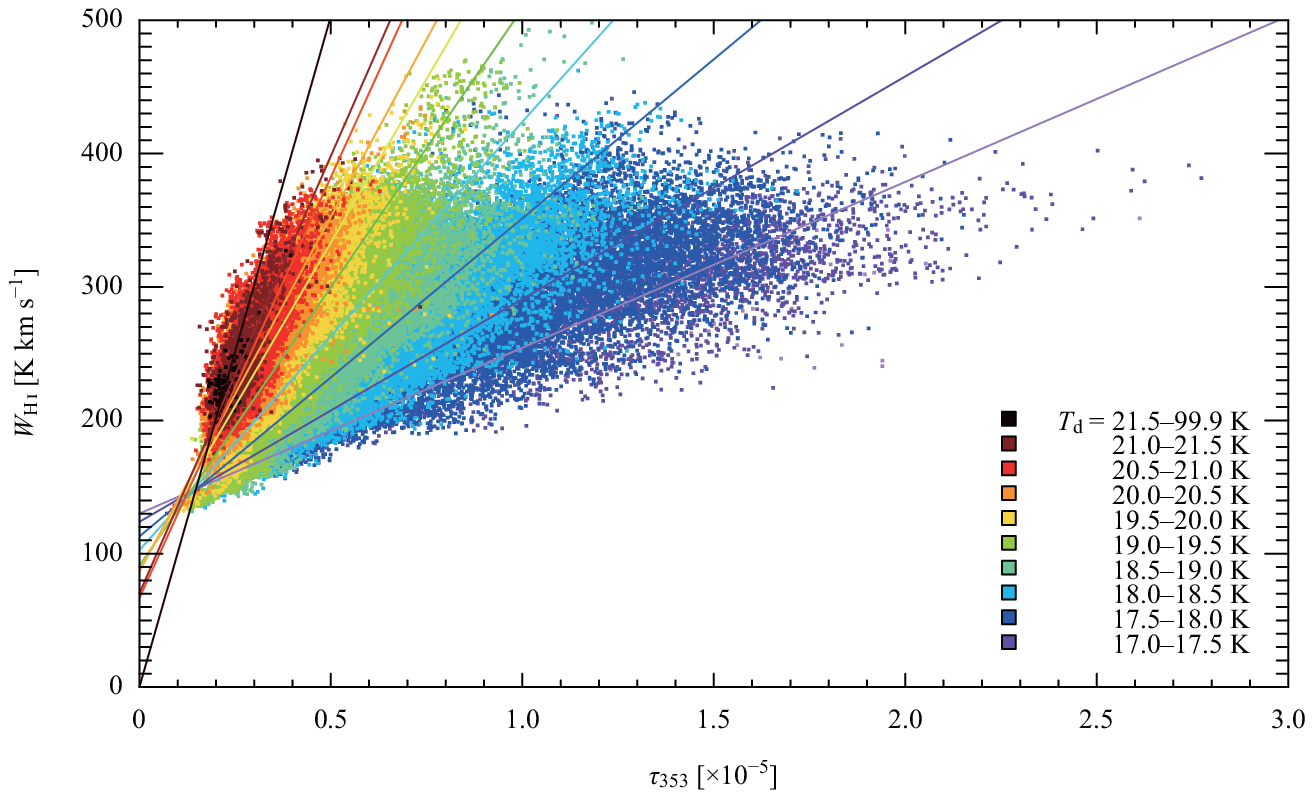}
    \caption{A $\taud$-$\WHI$ correlation plot in the MBM 53, 54, 55/HLCG 92{--}35 region \citep{2014ApJ...796...59F} colored by $\Td$. We used the {\Planck} dust data R1.20 unlike \citet{2014ApJ...796...59F}. The results of RMA (reduced major axis) regressions for each $\Td$ range are also plotted. We assumed that $\HI$ gas is optically-thin for the data points of $\Td\geq21.5\,\UK$ \citep{2014ApJ...796...59F}, therefore we make the regression line for $\Td\geq21.5\,\UK$ pass through the origin.}
    \label{fig:MBM535455_corr_tau353_WHI}
\end{figure*}

Figure~\ref{fig:MBM535455_corr_tau353_WHI} is a $\taud$-$\WHI$ correlation plot (similar to Figure~\ref{fig:Perseus_corr_tau353_WHI} in the present paper) for the {\MBM} and {\HLCG} region \citep{2014ApJ...796...59F,2003ApJ...592..217Y}. Although Figure~\ref{fig:MBM535455_corr_tau353_WHI} is almost the same as one in \citet{2014ApJ...796...59F}, the {\Planck} dust data R1.20 are used and we apply an additional mask which excludes the data points where intermediate-velocity clouds \citep[Pegasus-Pisces Arch,][]{2004ASSL..312...73A} are detected in Figure~\ref{fig:MBM535455_corr_tau353_WHI}. If we regard $\taud$ as a tracer of $\NH$ and if there is a positive correlation relationship between $\Td$ and $\Ts$, Figure~\ref{fig:Perseus_theoreticalcurves_NHI_WHI} and \ref{fig:MBM535455_corr_tau353_WHI} can roughly be considered as the same thing. Since it is thought that $\tauHI$ is larger if $\Td$ is lower, the variation width in $\WHI$ for the low-$\Td$ points is relatively large in Figure~\ref{fig:MBM535455_corr_tau353_WHI}, and vice versa. Accordingly, we performed linear fittings for each $\Td$ range and examined variances from each regression line. We used the reduced major axis (RMA) regression method \citep[e.g.,][]{1990ApJ...364..104I}, which minimizes the sum of the areas of the right-angled triangles delimited by each data point and the regression line, ${\sum}S_{i}$, and we defined the variance as the mean of the areas of the triangles, $\langle{S_{i}}\rangle$.

\begin{deluxetable}{cccc}
\tablewidth{\hsize}
\tablecaption{Relationship between $\Td$, $\langle{S_{i}}\rangle$, and $\langle\tauHI\rangle$\label{tab:Td_residual}}
\tablenum{3}
\tablehead{\colhead{} & \multicolumn{2}{c}{MBM 53, 54, 55} & \colhead{Perseus} \\
\colhead{$\Td$} & \colhead{$\langle{S_{i}}\rangle$} & \colhead{$\langle\tauHI\rangle$} & \colhead{$\langle{S_{i}}\rangle$} \\
\colhead{$[\UK]$} & \colhead{$[10^{-5}\,\UII]$} & \colhead{} & \colhead{$[10^{-5}\,\UII]$} \\
\colhead{(a)} & \colhead{(b)} & \colhead{(c)} & \colhead{(d)}}
\startdata
$21.5\leq$          & 0.34 & 0.09 & \nodata \\
$21.0\text{--}21.5$ & 0.34 & 0.14 & \nodata \\
$20.5\text{--}21.0$ & 0.45 & 0.22 & \nodata \\
$20.0\text{--}20.5$ & 0.48 & 0.30 & $\phn0.25$ \\
$19.5\text{--}20.0$ & 0.52 & 0.42 & $\phn0.37$ \\
$19.0\text{--}19.5$ & 0.76 & 0.66 & $\phn1.1\phn$ \\
$18.5\text{--}19.0$ & 1.09 & 1.01 & $\phn9.4\phn$ \\
$18.0\text{--}18.5$ & 1.63 & 1.54 & $11.9\phn$ \\
$17.5\text{--}18.0$ & 3.34 & 2.56 & $16.1\phn$ \\
$<17.5$             & 3.95 & 3.76 & $17.3\phn$ \\
\enddata
%\tablenotetext{(a)}{}
\tablecomments{The dispersions of the data points from the regression lines shown in Figure~\ref{fig:MBM535455_corr_tau353_WHI}. (a) The $\Td$ range. (b) and (c) The dispersions of the data points defined in the text, and the mean value of $\tauHI$ estimated in \citet{2014ApJ...796...59F}. We use the relationship between these two as a template. (d) The dispersions of the data points in the Perseus region.}
\end{deluxetable}

\begin{figure}[]
    \centering
    \includegraphics[scale=1]{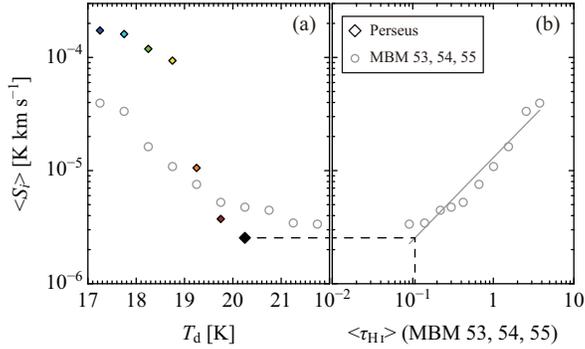}
    \caption{(a) Dispersions of the data points for each $\Td$ range. The open circles indicate the results for the MBM 53, 54, 55 region and we use this as a template. The diamonds indicate the result for the Perseus region. The colors of the diamonds are the same as Figure~\ref{fig:Perseus_corr_tau353_WHI_powerfit}. (b) The relationship between $\langle\tauHI\rangle$ \citep[average of $\tauHI$ derived by the same method as][]{2014ApJ...796...59F} and the variance for each $\Td$ range (MBM 53, 54, 55 template). The solid line indicates the result of a linear fit. The horizontal dashed line denotes $\langle{S_{i}}\rangle$ for the range of $\Td>20.0\,\UK$, and the vertical dashed line the corresponding $\langle\tauHI\rangle$ ($=0.11$).}
    \label{fig:Perseus_variance_template_MBM535455}
\end{figure}

The results are shown in Table~\ref{tab:Td_residual} and in Figure~\ref{fig:Perseus_variance_template_MBM535455}. In the MBM 53, 54, 55 region, an anti-correlation relationship between $\Td$ and $\langle{S_{i}}\rangle$ can obviously be seen (the columns 1 and 2 of Table~\ref{tab:Td_residual}, and Figure~\ref{fig:Perseus_variance_template_MBM535455}(a)). The column 3 of Table~\ref{tab:Td_residual} shows the mean $\tauHI$ values for each $\Td$ range derived by using the method described in \citet{2014ApJ...796...59F}, and we can see a positive correlation relationship against $\langle{S_{i}}\rangle$ (see also Figure~\ref{fig:Perseus_variance_template_MBM535455}(b)). Therefore, $\langle\tauHI\rangle$ for $\Td\geq20.0\,\UK$ for the Perseus region can be estimated by using the $\langle\tauHI\rangle${--}$\langle{S_{i}}\rangle$ relationship for the MBM 53, 54, 55 region as a template. The solid line in Figure~\ref{fig:Perseus_variance_template_MBM535455}(b) indicates the result of a linear fitting for the $\langle\tauHI\rangle${--}$\langle{S_{i}}\rangle$ relationship. By referring this line we get $\langle\tauHI\rangle$ for $\Td\geq20.0\,\UK$ for the Perseus region as $\langle\tauHI\rangle=0.11$ from $\langle{S_{i}}\rangle=0.5\times10^{-5}\,\UII$ (the dashed lines).

\begin{figure*}[]
    \centering
    \includegraphics[scale=1]{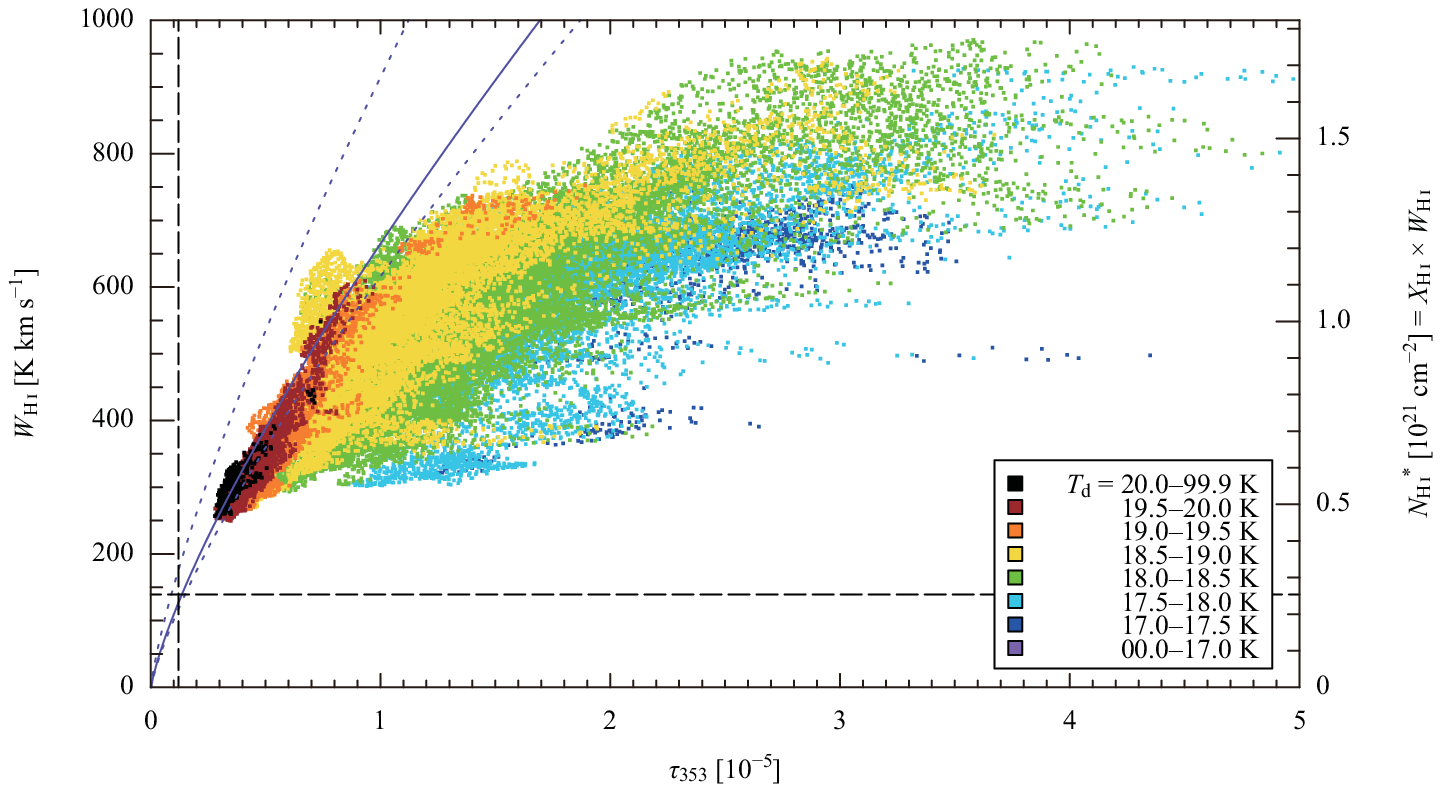}
    \caption{The $\taud$-$\WHI$ plot in the Perseus region. The right hand $y$-axis indicates the column number density of $\HI$ in the optically-thin limit ($\NHI$). Fitting the data points which $\Td>20.0\,\UK$ with the theoretical function fixing $\alpha=1.3$, and $\tauHI=0.11$, we get $\taudref=1.2\times10^{-6}$. Note that we use the relationship described in \citet{2015ApJ...798....6F}, $\NHref=(1.15\times10^{8})\times\XHI\times\taudref$. The left dotted curves indicates the case which $\taudref=4.77\times10^{-6}$, and the right one the case which $\taudref=8.7\times10^{-7}$. The dashed lines correspond to $\taudref=1.2\times10^{-6}$ and $\NHref=2.5\times10^{20}\,\UCND$.
    }
    \label{fig:Perseus_corr_tau353_WHI_powerfit}
\end{figure*}

By applying this $\langle\tauHI\rangle$ into Equation~(\ref{eq:curve_tau353_WHI_theoretical}) and with this function fitting the data points of $\Td\geq20.0\,\UK$, $\taudref$ for the Perseus region can be calculated. Note that $\alpha=1.3$ was applied and we used the relationship of $\NHref=(1.15\times10^{8})\times\XHI\times\taudref$ \citep{2015ApJ...798....6F}. The result of the fitting is plotted in Figure~\ref{fig:Perseus_corr_tau353_WHI_powerfit}, and we get $\taudref=1.2\times10^{-6}$ and $\NHref=2.5\times10^{20}\,\UCND$. The left-hand and right-hand dotted curves show the cases which $\taudref=4.77\times10^{-6}$ \citep{2015ApJ...798....6F}, and $\taudref=8.7\times10^{-7}$ \citep[corresponds to $\WHI=100\,\UII$ in][]{2015ApJ...798....6F}, respectively. Both of them cannot trace the distribution of the data points of $\Td\geq20.0\,\UK$.

From the above results, we adopted $\alpha=1.3$, $\taudref=1.2\times10^{-6}$, and $\NHref=2.5\times10^{20}\,\UCND$ for the Perseus region.

\bibliographystyle{aasjournal}
\bibliography{reference}

\begin{thebibliography}{}
\expandafter\ifx\csname natexlab\endcsname\relax\def\natexlab#1{#1}\fi
\providecommand{\url}[1]{\href{#1}{#1}}

\bibitem[{{Albert} \& {Danly}(2004)}]{2004ASSL..312...73A}
{Albert}, C.~E., \& {Danly}, L. 2004, in Astrophysics and Space Science
  Library, Vol. 312, High Velocity Clouds, ed. H.~{van Woerden}, B.~P.
  {Wakker}, U.~J. {Schwarz}, \& K.~S. {de Boer}, 73

\bibitem[{{Bally} {et~al.}(2008){Bally}, {Walawender}, {Johnstone}, {Kirk}, \&
  {Goodman}}]{2008hsf1.book..308B}
{Bally}, J., {Walawender}, J., {Johnstone}, D., {Kirk}, H., \& {Goodman}, A.
  2008, {The Perseus Cloud}, ed. B.~{Reipurth}, 308

\bibitem[{{Bolatto} {et~al.}(2013){Bolatto}, {Wolfire}, \&
  {Leroy}}]{2013ARA&A..51..207B}
{Bolatto}, A.~D., {Wolfire}, M., \& {Leroy}, A.~K. 2013, \araa, 51, 207

\bibitem[{{Cotten} \& {Magnani}(2013)}]{2013MNRAS.436.1152C}
{Cotten}, D.~L., \& {Magnani}, L. 2013, \mnras, 436, 1152

\bibitem[{{Dame} {et~al.}(2001){Dame}, {Hartmann}, \&
  {Thaddeus}}]{2001ApJ...547..792D}
{Dame}, T.~M., {Hartmann}, D., \& {Thaddeus}, P. 2001, \apj, 547, 792

\bibitem[{{Dickey} \& {Lockman}(1990)}]{1990ARA&A..28..215D}
{Dickey}, J.~M., \& {Lockman}, F.~J. 1990, \araa, 28, 215

\bibitem[{{Draine}(2011)}]{2011piim.book.....D}
{Draine}, B.~T. 2011, {Physics of the Interstellar and Intergalactic Medium}

\bibitem[{{Finkbeiner}(2003)}]{2003ApJS..146..407F}
{Finkbeiner}, D.~P. 2003, \apjs, 146, 407

\bibitem[{{Forbrich} {et~al.}(2015){Forbrich}, {Lada}, {Lombardi},
  {Rom{\'a}n-Z{\'u}{\~n}iga}, \& {Alves}}]{2015A&A...580A.114F}
{Forbrich}, J., {Lada}, C.~J., {Lombardi}, M., {Rom{\'a}n-Z{\'u}{\~n}iga}, C.,
  \& {Alves}, J. 2015, \aap, 580, A114

\bibitem[{{Fukui} {et~al.}(2017){Fukui}, {Hayakawa}, {Inoue}, {Torii},
  {Okamoto}, {Tachihara}, {Onishi}, \& {Hayashi}}]{2017arXiv170107129F}
{Fukui}, Y., {Hayakawa}, T., {Inoue}, T., {et~al.} 2017, ArXiv e-prints,
  arXiv:1701.07129

\bibitem[{{Fukui} {et~al.}(2015){Fukui}, {Torii}, {Onishi}, {Yamamoto},
  {Okamoto}, {Hayakawa}, {Tachihara}, \& {Sano}}]{2015ApJ...798....6F}
{Fukui}, Y., {Torii}, K., {Onishi}, T., {et~al.} 2015, \apj, 798, 6

\bibitem[{{Fukui} {et~al.}(2012){Fukui}, {Sano}, {Sato}, {Torii}, {Horachi},
  {Hayakawa}, {McClure-Griffiths}, {Rowell}, {Inoue}, {Inutsuka}, {Kawamura},
  {Yamamoto}, {Okuda}, {Mizuno}, {Onishi}, {Mizuno}, \&
  {Ogawa}}]{2012ApJ...746...82F}
{Fukui}, Y., {Sano}, H., {Sato}, J., {et~al.} 2012, \apj, 746, 82

\bibitem[{{Fukui} {et~al.}(2014){Fukui}, {Okamoto}, {Kaji}, {Yamamoto},
  {Torii}, {Hayakawa}, {Tachihara}, {Dickey}, {Okuda}, {Ohama}, {Kuroda}, \&
  {Kuwahara}}]{2014ApJ...796...59F}
{Fukui}, Y., {Okamoto}, R., {Kaji}, R., {et~al.} 2014, \apj, 796, 59

\bibitem[{{Gillmon} {et~al.}(2006){Gillmon}, {Shull}, {Tumlinson}, \&
  {Danforth}}]{2006ApJ...636..891G}
{Gillmon}, K., {Shull}, J.~M., {Tumlinson}, J., \& {Danforth}, C. 2006, \apj,
  636, 891

\bibitem[{{G{\'o}rski} {et~al.}(2005){G{\'o}rski}, {Hivon}, {Banday},
  {Wandelt}, {Hansen}, {Reinecke}, \& {Bartelmann}}]{2005ApJ...622..759G}
{G{\'o}rski}, K.~M., {Hivon}, E., {Banday}, A.~J., {et~al.} 2005, \apj, 622,
  759

\bibitem[{{Heiles} \& {Troland}(2003)}]{2003ApJS..145..329H}
{Heiles}, C., \& {Troland}, T.~H. 2003, \apjs, 145, 329

\bibitem[{{Inoue} \& {Inutsuka}(2012)}]{2012ApJ...759...35I}
{Inoue}, T., \& {Inutsuka}, S.-i. 2012, \apj, 759, 35

\bibitem[{{Isobe} {et~al.}(1990){Isobe}, {Feigelson}, {Akritas}, \&
  {Babu}}]{1990ApJ...364..104I}
{Isobe}, T., {Feigelson}, E.~D., {Akritas}, M.~G., \& {Babu}, G.~J. 1990, \apj,
  364, 104

\bibitem[{{Jones} {et~al.}(2013){Jones}, {Fanciullo}, {K{\"o}hler},
  {Verstraete}, {Guillet}, {Bocchio}, \& {Ysard}}]{2013A&A...558A..62J}
{Jones}, A.~P., {Fanciullo}, L., {K{\"o}hler}, M., {et~al.} 2013, \aap, 558,
  A62

\bibitem[{{Juvela} \& {Montillaud}(2016)}]{2016AaA...585A..38J}
{Juvela}, M., \& {Montillaud}, J. 2016, \aap, 585, A38

\bibitem[{{Landsman}(1993)}]{1993ASPC...52..246L}
{Landsman}, W.~B. 1993, in Astronomical Society of the Pacific Conference
  Series, Vol.~52, Astronomical Data Analysis Software and Systems II, ed.
  R.~J. {Hanisch}, R.~J.~V. {Brissenden}, \& J.~{Barnes}, 246

\bibitem[{{Langer} {et~al.}(2014){Langer}, {Velusamy}, {Pineda}, {Willacy}, \&
  {Goldsmith}}]{2014A&A...561A.122L}
{Langer}, W.~D., {Velusamy}, T., {Pineda}, J.~L., {Willacy}, K., \&
  {Goldsmith}, P.~F. 2014, \aap, 561, A122

\bibitem[{{Lee} {et~al.}(2014){Lee}, {Stanimirovi{\'c}}, {Wolfire}, {Shetty},
  {Glover}, {Molina}, \& {Klessen}}]{2014ApJ...784...80L}
{Lee}, M.-Y., {Stanimirovi{\'c}}, S., {Wolfire}, M.~G., {et~al.} 2014, \apj,
  784, 80

\bibitem[{{Lee} {et~al.}(2012){Lee}, {Stanimirovi{\'c}}, {Douglas}, {Knee}, {Di
  Francesco}, {Gibson}, {Begum}, {Grcevich}, {Heiles}, {Korpela}, {Leroy},
  {Peek}, {Pingel}, {Putman}, \& {Saul}}]{2012ApJ...748...75L}
{Lee}, M.-Y., {Stanimirovi{\'c}}, S., {Douglas}, K.~A., {et~al.} 2012, \apj,
  748, 75

\bibitem[{{Lombardi} \& {Alves}(2001)}]{2001AaA...377.1023L}
{Lombardi}, M., \& {Alves}, J. 2001, \aap, 377, 1023

\bibitem[{{Magnani} {et~al.}(1998){Magnani}, {Onello}, {Adams}, {Hartmann}, \&
  {Thaddeus}}]{1998ApJ...504..290M}
{Magnani}, L., {Onello}, J.~S., {Adams}, N.~G., {Hartmann}, D., \& {Thaddeus},
  P. 1998, \apj, 504, 290

\bibitem[{{McClure-Griffiths} {et~al.}(2006){McClure-Griffiths}, {Dickey},
  {Gaensler}, {Green}, \& {Haverkorn}}]{2006ApJ...652.1339M}
{McClure-Griffiths}, N.~M., {Dickey}, J.~M., {Gaensler}, B.~M., {Green}, A.~J.,
  \& {Haverkorn}, M. 2006, \apj, 652, 1339

\bibitem[{{Miville-Desch{\^e}nes} \& {Lagache}(2005)}]{2005ApJS..157..302M}
{Miville-Desch{\^e}nes}, M.-A., \& {Lagache}, G. 2005, \apjs, 157, 302

\bibitem[{{Peek} {et~al.}(2011){Peek}, {Heiles}, {Douglas}, {Lee}, {Grcevich},
  {Stanimirovi{\'c}}, {Putman}, {Korpela}, {Gibson}, {Begum}, {Saul},
  {Robishaw}, \& {Kr{\v c}o}}]{2011ApJS..194...20P}
{Peek}, J.~E.~G., {Heiles}, C., {Douglas}, K.~A., {et~al.} 2011, \apjs, 194, 20

\bibitem[{{\textit{Planck} Collaboration}
  {et~al.}(2011{\natexlab{a}}){\textit{Planck} Collaboration}, {Ade},
  {Aghanim}, {Arnaud}, {Ashdown}, {Aumont}, {Baccigalupi}, {Baker}, {Balbi},
  {Banday}, \& et~al.}]{2011A&A...536A...1P}
{\textit{Planck} Collaboration}, {Ade}, P.~A.~R., {Aghanim}, N., {et~al.}
  2011{\natexlab{a}}, \aap, 536, A1

\bibitem[{{\textit{Planck} Collaboration}
  {et~al.}(2011{\natexlab{b}}){\textit{Planck} Collaboration}, {Ade},
  {Aghanim}, {Arnaud}, {Ashdown}, {Aumont}, {Baccigalupi}, {Balbi}, {Banday},
  {Barreiro}, \& et~al.}]{2011A&A...536A..19P}
---. 2011{\natexlab{b}}, \aap, 536, A19

\bibitem[{{\textit{Planck} Collaboration}
  {et~al.}(2011{\natexlab{c}}){\textit{Planck} Collaboration}, {Abergel},
  {Ade}, {Aghanim}, {Arnaud}, {Ashdown}, {Aumont}, {Baccigalupi}, {Balbi},
  {Banday}, \& et~al.}]{2011AaA...536A..24P}
{\textit{Planck} Collaboration}, {Abergel}, A., {Ade}, P.~A.~R., {et~al.}
  2011{\natexlab{c}}, \aap, 536, A24

\bibitem[{{\textit{Planck} Collaboration} {et~al.}(2014){\textit{Planck}
  Collaboration}, {Abergel}, {Ade}, {Aghanim}, {Alves}, {Aniano},
  {Armitage-Caplan}, {Arnaud}, {Ashdown}, {Atrio-Barandela}, \&
  et~al.}]{2014AaA...571A..11P}
---. 2014, \aap, 571, A11

\bibitem[{{Rachford} {et~al.}(2002){Rachford}, {Snow}, {Tumlinson}, {Shull},
  {Blair}, {Ferlet}, {Friedman}, {Gry}, {Jenkins}, {Morton}, {Savage},
  {Sonnentrucker}, {Vidal-Madjar}, {Welty}, \& {York}}]{2002ApJ...577..221R}
{Rachford}, B.~L., {Snow}, T.~P., {Tumlinson}, J., {et~al.} 2002, \apj, 577,
  221

\bibitem[{{Reich} \& {Reich}(1986)}]{1986AaAS...63..205R}
{Reich}, P., \& {Reich}, W. 1986, \aaps, 63, 205

\bibitem[{{Reich}(1982)}]{1982AaAS...48..219R}
{Reich}, W. 1982, \aaps, 48, 219

\bibitem[{{Ridge} {et~al.}(2006){Ridge}, {Schnee}, {Goodman}, \&
  {Foster}}]{2006ApJ...643..932R}
{Ridge}, N.~A., {Schnee}, S.~L., {Goodman}, A.~A., \& {Foster}, J.~B. 2006,
  \apj, 643, 932

\bibitem[{{Roy} {et~al.}(2013){Roy}, {Martin}, {Polychroni}, {Bontemps},
  {Abergel}, {Andr{\'e}}, {Arzoumanian}, {Di Francesco}, {Hill}, {Konyves},
  {Nguyen-Luong}, {Pezzuto}, {Schneider}, {Testi}, \&
  {White}}]{2013ApJ...763...55R}
{Roy}, A., {Martin}, P.~G., {Polychroni}, D., {et~al.} 2013, \apj, 763, 55

\bibitem[{{Schultheis} {et~al.}(2014){Schultheis}, {Chen}, {Jiang}, {Gonzalez},
  {Enokiya}, {Fukui}, {Torii}, {Rejkuba}, \& {Minniti}}]{2014A&A...566A.120S}
{Schultheis}, M., {Chen}, B.~Q., {Jiang}, B.~W., {et~al.} 2014, \aap, 566, A120

\bibitem[{{Sofue}(2013)}]{2013PASJ...65..118S}
{Sofue}, Y. 2013, \pasj, 65, 118

\bibitem[{{Stanimirovi{\'c}} {et~al.}(2014){Stanimirovi{\'c}}, {Murray}, {Lee},
  {Heiles}, \& {Miller}}]{2014ApJ...793..132S}
{Stanimirovi{\'c}}, S., {Murray}, C.~E., {Lee}, M.-Y., {Heiles}, C., \&
  {Miller}, J. 2014, \apj, 793, 132

\bibitem[{{T{\'o}th} {et~al.}(2014){T{\'o}th}, {Marton}, {Zahorecz},
  {Bal{\'a}zs}, {Ueno}, {Tamura}, {Kawamura}, {Kiss}, \&
  {Kitamura}}]{2014PASJ...66...17T}
{T{\'o}th}, L.~V., {Marton}, G., {Zahorecz}, S., {et~al.} 2014, \pasj, 66, 17

\bibitem[{{Wenger} {et~al.}(2000){Wenger}, {Ochsenbein}, {Egret}, {Dubois},
  {Bonnarel}, {Borde}, {Genova}, {Jasniewicz}, {Lalo{\"e}}, {Lesteven}, \&
  {Monier}}]{2000AaAS..143....9W}
{Wenger}, M., {Ochsenbein}, F., {Egret}, D., {et~al.} 2000, \aaps, 143, 9

\bibitem[{{Wolfire} {et~al.}(2010){Wolfire}, {Hollenbach}, \&
  {McKee}}]{2010ApJ...716.1191W}
{Wolfire}, M.~G., {Hollenbach}, D., \& {McKee}, C.~F. 2010, \apj, 716, 1191

\bibitem[{{Yamamoto} {et~al.}(2003){Yamamoto}, {Onishi}, {Mizuno}, \&
  {Fukui}}]{2003ApJ...592..217Y}
{Yamamoto}, H., {Onishi}, T., {Mizuno}, A., \& {Fukui}, Y. 2003, \apj, 592, 217

\bibitem[{{Ysard} {et~al.}(2015){Ysard}, {K{\"o}hler}, {Jones},
  {Miville-Desch{\^e}nes}, {Abergel}, \& {Fanciullo}}]{2015A&A...577A.110Y}
{Ysard}, N., {K{\"o}hler}, M., {Jones}, A., {et~al.} 2015, \aap, 577, A110

\end{thebibliography}

\end{document}